# On the Magnetic Current Density in Maxwell's Equations Based on Noether's Theorem

Mehdi Nosrati

*McGill University, Montreal, Quebec, H3A 0E9, Canada*

**Abstract**

**Despite the search for supersymmetry based on abelian and non-abelian Yang-Mills gauge field theory, the Maxwell equations, as the earliest gauge field theory, are non-symmetric because of the undefined term of magnetic current density. This article reports on the theoretical quantization of this term based on spontaneous symmetry breaking in the spatial geometry of a gauge group (G-group) of quantum charged (QC) particles. A locally supersymmetric background-independent spatial geometry of the G-group is developed based on the commutative string field interaction (SFI) between infinite number of QC particles and the Dirac-'t Hooft-Polyakov grand monopole and the noncommutative SFI of each pair of adjacent QC particles in the G-group. Two adjoint and disjoint currents are associated with the commutative and noncommutative SFIs, respectively, based on the spin of a QC particle. The adjoint and disjoint currents are associated with a zero resistivity between a QC particle and the grand monopole and an infinite resistivity between each pair of adjacent QC particles correlated to their microscopic commutative and noncommutative SFIs in the G-group, respectively. This article demonstrates that the two corresponding resistivities are finite (greater than zero and less than infinity) for the macroscopic commutative and noncommutative SFIs of the G-group. Therefore, the adjoint and disjoint currents are related to the classical macroscopic currents known as electric and magnetic currents, respectively. Because the microscopic adjoint current associated with the commutative SFI has already been related to the macroscopic electric current density in Maxwell's equations, it is proposed that the microscopic disjoint current associated with the noncommutative SFI is related to the undefined magnetic current density in these equations.**

**Keywords:** Magnetic current density, supersymmetric background-independent spatial geometry of gauge group of QC particles, adjoin and disjoint currents, commutative and noncommutative string field interactions.

## 1. Introduction

Maxwell's equations unify the Gauss, Ampere, and Faraday laws, which are based on experimental observations of macroscopic electric and magnetic fields associated with charge and current [1]. Despite the search for supersymmetry based on abelian and non-abelian gauge theories [2], Maxwell's equations, as the earliest gauge field theory, remain incomplete: the undefined term of magnetic current density has not been theoretically explained but is instead presented as a fictitious parameter. The unexplained term of magnetic current density has been associated with a magnetic monopole; however, this term requires further verification [3]-[7].

Maxwell's equations were classically derived based on the exterior macroscopic field interactions of a dipole magnet as the explicit source of an electromagnetic field [1]. However, by generalizing the principle of QCD particle confinement [8]-[9] to QED particles (electrons) in the most exterior shells of an atomic structure, the magnetic current term associated with the magnetic monopole can be elucidated by analyzing the interior microscopic field interactions in the atomic structure of a dipole magnet.

An interior field analysis of the atomic structure of a dipole magnet as a gauge group of QC (QCD and QED) particles necessitates the compactification of both commutative and noncommutative string field interactions (SFIs) in a single gauge group of QC particles to represent all fundamental interactions and to demonstrate the ultimate building block predicted by Steven Weinberg and Gerard 't Hooft [10]-[13].

The existence of a monopole was first posited by Paul Dirac, based on the quantization of electric charges [3]. 't Hooft and Polyakov theoretically studied the SFIs of a magnetic monopole with respect to other fundamental interactions and demonstrated the necessity of a magnetic monopole in the grand unified gauge theory [4] and [6]. The Dirac-'t Hooft- Polykov grand monopole, as the core of the atomic structure, is predicted to carry magnetic and color charges and to interact with both QED particles under the $U(1)_{EM}$ gauge group and QCD particles under the $SU(2)$ and $SU(3)$ gauge groups associated with the electroweak and strong forces (Figure 1). Despite the profound influence of this work on the unification of the magnetic monopole with other fundamental interactions, the magnetic current density term in Maxwell's equations remains unexplained.

This study presents a theoretical explanation of the magnetic current density, aiming to symmetrize Maxwell's equations as the earliest gauge field equations. The unexplained magnetic current density is shown to be crucial in clarifying the relationships among the fundamental interactions. By considering a gauge group (G-group) of QC particles, the undefined magnetic current term is shown to depend on several physical phenomena, including the spin of QC particles and their corresponding SFIs in a G-group [14], the spatial geometry of the G-group [15], supersymmetry breaking of the spatial geometry of the G-group [16]-[18], and ultimately the resistance of the G-group against supersymmetry breaking due to non-abelian SFIs of the G-group with external string fields [19].

To mathematically derive the magnetic current density, mutual correlations among the related physical phenomena are studied. A supersymmetric background-independent G-group of QC particles is considered on the basis of 1/N expansion theory [20]-[21], in which N QC particles are confined, as the elements of the G-group, with the grand monopole as the identity element of the G-group. The spatial geometry of the G-group is developed from the commutative SFIs of N QC particles with the grand monopole and the noncommutative SFIs of each pair of adjacent QC particles in the G-group.

To physically distinguish the commutative and noncommutative SFIs in the G-group, the corresponding adjoint and disjoint currents will be geometrically explored based on the spin of a QC particle. The two conjugated Dirac field functions are associated with the string fields of each QC particle and the grand monopole in the G-group, mathematically representing the commutative SFI. This representation describes the inherently identical string fields of the QC particle and the grand monopole, which are distinguished by opposite directions with respect to the grand monopole as the point of reference (inward and/or outward directions of the vector string fields for the grand monopole and QC particles; see Figure 2(a). In contrast, the noncommutative SFI between each pair of adjacent QC particles in the 2D spatial geometry of the G-group is mathematically illustrated by a pair of nonconjugated field functions. These representations are shown to address the physical origin of the non-integrable constant phase suggested by Dirac for the mathematical treatment of singularities in an electromagnetic field [22], while also addressing the compactification of both commutative and noncommutative Yang-Mills SFIs in the spatial geometry of a single G-group of QC particles [10].

By including both commutative and noncommutative SFIs in the supersymmetric G-group of QC particles, the disjoint current is shown to momentarily intensify due to spontaneous supersymmetry breaking in the spatial geometry of the G-group under interactions with the string fields of another G-group of QC particles (external B-field in [10]).

The adjoint and disjoint currents and their corresponding commutative and noncommutative SFIs are mathematically explored on the basis of Hermitian and anti-Hermitian (skew-Hermitian) models and the corresponding conjugated and nonconjugated string field functions. The adjoint and disjoint currents are shown to be dynamically exchanged in a G-group of QC particles. The adjoint current is associated with the macroscopic electric current density in electromagnetic field theory, and it is proposed that the disjoint current is associated with the undefined term of macroscopic magnetic current density in Maxwell's equations.

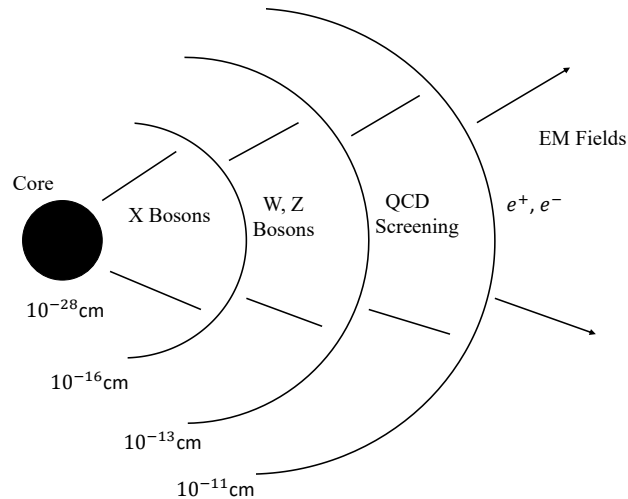

Figure 1. Grand monople unified with the fundamental field interactions [6].

## 2. Review: Classical non-relativistic commutative and noncommutative SFIs of like- and unlike-charged particles

According to Wilson, Mandelstam, 't Hooft, Polykov, and Nambu researches [23]-[24], it has been concluded that the lines of charges are confined because they are connected to quantum particles through strings of energy. This primary conclusion did not comply with non-abelian gauge field theory, thus, Polykov conjectured the existence of a duality between gauge fields and string. The key point of Polykov's work is the development of Wilson loop action on the basis of monopole configuration [23]. Consideration of monopole constrain, compliance with abelian and non-abelian Yang-Mills gauge field theory, the undefined magnetic current density in Maxwell's equations, and the dynamic exchange of closed and open strings, a line of charge is hypothesized to be permanently connected to a quantum particle at one end (Dirichlet boundary condition [25]) and temporarily confined between the quantum particle and its anti-particle (unlike charged particles). This hypothesis adds a degree of freedom by which the confined string fields can be momentarily free at the other end in the case of SFIs, realization of open string fields with Dirichlet-Neuman boundary conditions[10] and [25] (see Section 5, Figure 8 and Figure 9).

Figure 2(a) shows two quantum charged (QC) particles (QED or/and QCD particles) with corresponding type I strings [24]. The non-relativistic SFIs of the two sets of like- and unlike-charged quantum particles can be distinguished according to the intrinsic string field couplings between the particles [26]. As shown in Figure 2(b), a noncommutative SFI (coupling) occurs between the string fields of two like-charged quantum particles when an external force brings the two QC particles close to each other (forced position of the particles). In contrast, a commutative SFI is demonstrated between the string fields of two unlike-charged quantum particles in their natural positions (Figure 2(c)) [27].

The commutative and noncommutative SFIs can be simply exemplified by the charge-emitting and absorbing phenomena between a pair of QC particles, following the concept of resistivity. By taking the two distinguished states, closed and open, of a string between two particles in string theory [28], one can simply state that no closed string field is realized between two like-charged quantum particles (Dirichlet-Nauman boundary conditions for the string fields of the QC particles [25]). This statement implies that the string fields of each particle encounter an infinite resistance to the construction of zero-mode closed strings with the other particle.

However, the string fields of unlike-charged quantum particles interact in such a way that closed string fields are confined between the particles, with Dirichlet-Dirichlet boundary conditions for the strings. In contrast to the first case, the string fields of each particle encounter a zero resistance to the formation of zero-mode closed strings with the other particle. Therefore, the field analysis of each particle pair provides two discrete resistances of zero and/or infinity, depending on the nature of the particle charges or the inward and/or outward directions of their corresponding string fields.

The literature on quantum particle theory has primarily focused on abelian SFIs and equations of motion based on conjugated Dirac field functions and corresponding Hermitian adjoint operator associated with a confined particle-antiparticle pair [8] and [29].

Considering the unlike-charged quantum particle-antiparticle confinement principle, the resistance encountered by the string fields of a QC particle for a commutative SFI (closed strings) would inherently be zero. As will be discussed below, the concept of resistance is invoked when the compactification of both the commutative and noncommutative SFIs with resistances of zero and infinity are studied in the spatial geometry of a single G-group of QC particles on the basis of abelian and non-abelian field theories. In relation to classical physics, the commutative and noncommutative SFIs can be associated with attractive and repulsive forces that produce Hermitian and anti-Hermitian operators. By considering these dual actions for electromagnetic gauge field theory, as the border between quantum mechanics (Planck scale) and general relativity (cosmological scale), some controversial concepts can be explained in the unification of quantum mechanics and general relativity, e.g., the repulsive cosmological force [30] and/or the link between positive and negative cosmological constants predicted by anti-de Sitter-de Sitter conformal field theory (AdS/dS CFT) correspondence [31].

Before geometrically illustrating the spin of a QC particle, specifically an electron, based on the commutative and noncommutative SFIs of the QC particle, a fundamental concept should be clarified. Dirac suggested adding a non-integrable constant to the general phase of a QC particle wave function [22], stating that this definite phase value is not a function of the spacetime, but rather a background-independent phase. This constant phase was suggested for the mathematical treatment of electromagnetic wave singularities at a phase of $\theta = \pi$. However, the physical origin of the non-integrable background-independent phase has not been explained. Recently, the concept of a background-independent approach, which originated from Dirac's suggestion, has been raised [32]-[34].

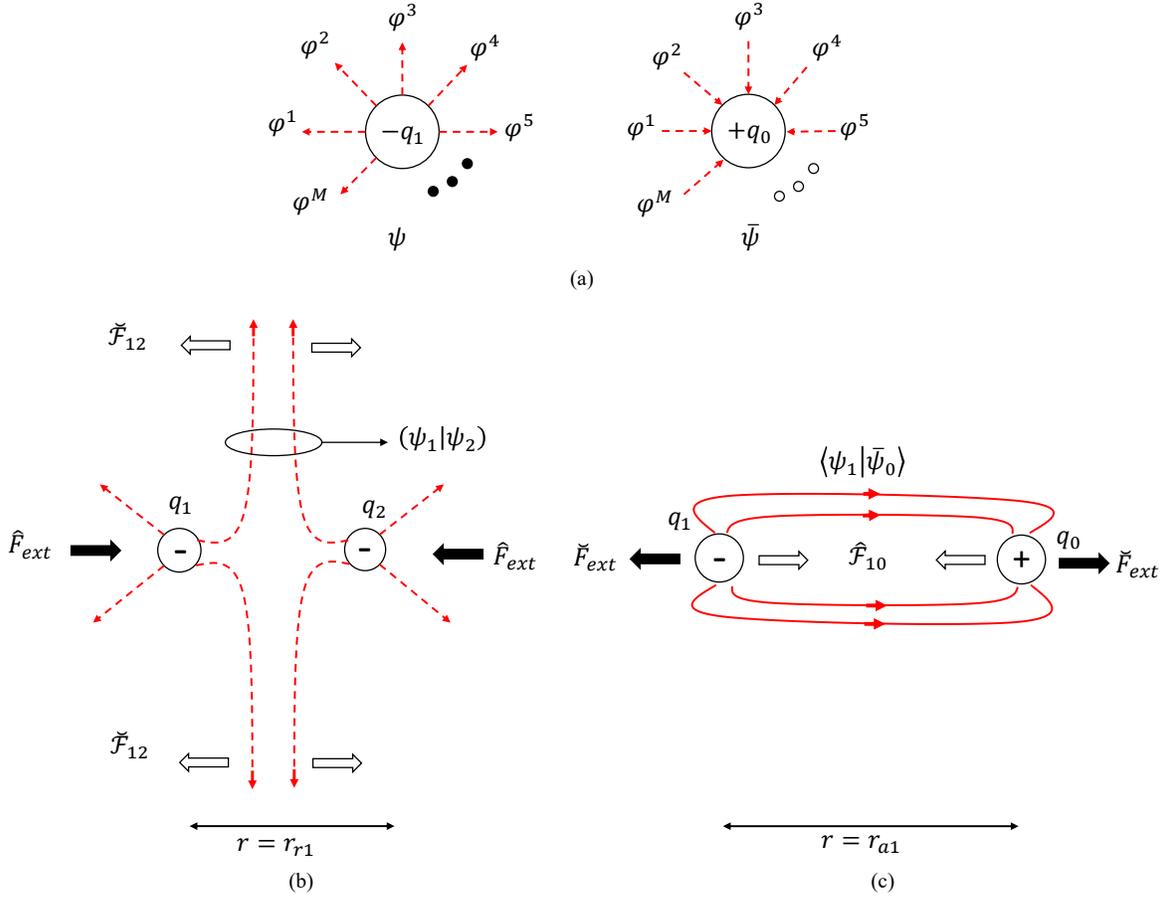

Figure 2. (a) Two unlike-charged particles (b) like-charged particles with noncommutative SFI (c) unlike-charged particles with commutative SFI [27] (The dashed and sloid lines are used to distinguish the noncommutative (open string) and commutative (closed string) SFIs, a string field is assumed to be a continuous entity).

In this work, the non-integrable background-independent constant phase identified by Dirac is initially clarified based on the vector background-independent string field directions of two QC particles with respect to each other and is later used to develop the spatial geometry of a G-group of QC particles. The clarification of this non-integrable background-independent constant phase provides the following contributions:

- Distinguishes the element of a G-group from the identity element of the G-group according to the different corresponding string field behaviors (vector background-independent directions). In this study, the Dirac-'t Hooft-Polyakov grand monopole is considered as the identity, and $N$ ($N \to \infty$) QC particles are considered as the elements of the G-group (see Section 4).
- Characterizes the commutative and noncommutative SFIs and compactifies them in the spatial geometry of the G-group of QC particles.
- Individualizes the microscopic adjoint and disjoint currents (Majorana supercurrent [35]), which are associated with the commutative and noncommutative SFIs as well as the macroscopic electric and magnetic currents in Maxwell's equations.
- Addresses a dynamically fixed background characterized by the non-relativistic constant phases of string fields of QC particles and the variable relativistic phases of wave functions associated with the QC particles. A background-independent constant phase is associated with the confined string fields of a QC particle in the spatial geometry of its corresponding G-group [23], while a background-dependent (relativistic) variable phase is associated with the perturbative variations of the string fields of the QC particle or its corresponding G-group under SFIs with external string fields (see Section 8.3).

The negatively and positively charged particles and the corresponding quantized string fields shown in Figure 2(a) are used to explain the non-integrable constant phase based on the intrinsic background-independent string field phase difference of the charged particles. The two particles can be fundamentally distinguished according to the vector directions of their string fields by taking one of the particles as a point of reference. By assessing the inward and/or outward vector directions of the particle string fields, a constant phase difference of $\pi$ is considered for the string fields of the particles, independent of Minkowski spacetime.

This background-independent phase difference is mathematically realized by assigning two conjugate Dirac string field functions for the positively and negatively charged quantum particles. The intrinsic constant phase difference of $\pi$ between two conjugated imaginary numbers can uniquely satisfy this condition. By considering quanta $M(M \to \infty)$ string fields associated with each particle in Figure 2(a), the string field functions of the two charged particles can be represented as follows:

$$\psi_{q_1}(x^\mu) = e^{-i\phi_{q_1}} \sum_{m=1}^{M} \varphi_{q_1}^m(x^\mu) e^{\pm j\phi_{q_1}^m(x^\mu)} = e^{-i\phi_{q_1}} \left(\psi_{q_1}(x^\mu)\right) \tag{2.1a}$$

$$\bar{\psi}_{q_0}(x^\mu) = e^{+i\phi_{q_0}} \sum_{m=1}^{M} \varphi_{q_0}^m(x^\mu) e^{\pm j\phi_{q_0}^m(x^\mu)} = e^{+i\phi_{q_0}} \left(\psi_{q_0}(x^\mu)\right) \tag{2.1b}$$

The $\{\phi_{q_1}, \phi_{q_0}\}$ and $\{\phi_{q_1}^m, \phi_{q_0}^m\}$ phases are assigned to the vector string fields of the particles to characterize both their background-independent (fixed) and background-dependent (relativistic-dynamic) string field functions.

These string field functions are used to determine the adjoint current between two unlike-charged quantum particles associated with the commutative SFI. In contrast, two nonconjugated string field functions are considered to derive the disjoint current between two like-charged quantum particles, in association with the noncommutative SFI.

The mathematical expressions of the string field functions of a pair of unlike-charged quantum particles cancel each other out because of the background-dependent intrinsic phase difference of $\pi$. From a physical point of view, the non-observable string fields are confined between the two unlike-charged quantum particles as rest string fields or vector potential string fields [23]. In contrast, the mathematical expressions of the string field functions of two like-charged quantum particles do not cancel each other out because they have a background-dependent intrinsic phase difference of zero. From a physical point of view, the observable string fields are curved between the two like-charged quantum particles as moving string fields or vector kinetic string fields.

## 3. Spin of a QC particle; adjoint and disjoint currents

The electron spin is an underlying principle in quantum mechanics derived from experimental measurements [36]. Although this principle has been mathematically characterized [14], some controversies remain [37]. Here, the aim is to relate the spin of a QC particle (electron) to other fundamental principles in the literature.

The distinct down and up states of the spin of an electron, with positive and negative half-integers, are correlated with the background-independent intrinsic string field phases of the like- and unlike-charged quantum particles discussed in Section 2. The spin of an electron's string fields is associated with the commutative and noncommutative SFIs. The second type of microscopic current is deduced and put forward with the nomenclature of disjoint current, similar to the microscopic adjoint current, corresponding to noncommutative and commutative SFIs, respectively.

The SFIs of an electron and two Hermitian (Dirac) conjugated string field functions, associated with the macroscopic electromagnetic fields of a dipole magnet, are theoretically studied in an abstract Hilbert space to identify the relationship between the spin of an electron's string fields and the conventional adjoint and newly suggested disjoint current.

The electron string fields are shown to break the adjoint geometrical symmetry of the Hermitian conjugated string fields of the dipole magnet due to the noncommutative SFI [38]. The new locally nonsymmetrical geometry includes both disjoint and adjoint currents corresponding to the noncommutative and commutative SFIs. It will be shown later that this abstract locally nonsymmetrical geometry contributes to the global supersymmetry of a G-group of QC particles.

Ultimately, the adjoint $\hat{j}^\mu(x)$ and disjoint $m_\mu(x)$ currents are postulated as microscopic currents associated with the macroscopic electric and magnetic current densities in Maxwell's equations. It is postulated that the noncommutative SFI generates a disjoint current corresponding to the magnetic field of $\mathcal{H}$ as the duality of the adjoint current corresponding to the electric field of $\mathcal{E}$ in Maxwell's equations.

As shown in Figure 3, an electron with a total charge of $q_e$ and a corresponding string field function of $\psi_{q_e}$ is assumed to be enclosed by barrier B, indicating that there is no initial SFI between the electron and the dipole magnet. The conventional conserved Noether's current is initially reviewed for the commutative SFI of the two poles of the dipole magnet, and then, the scenario is evaluated for the SFIs of the electron and the dipole magnet [39].

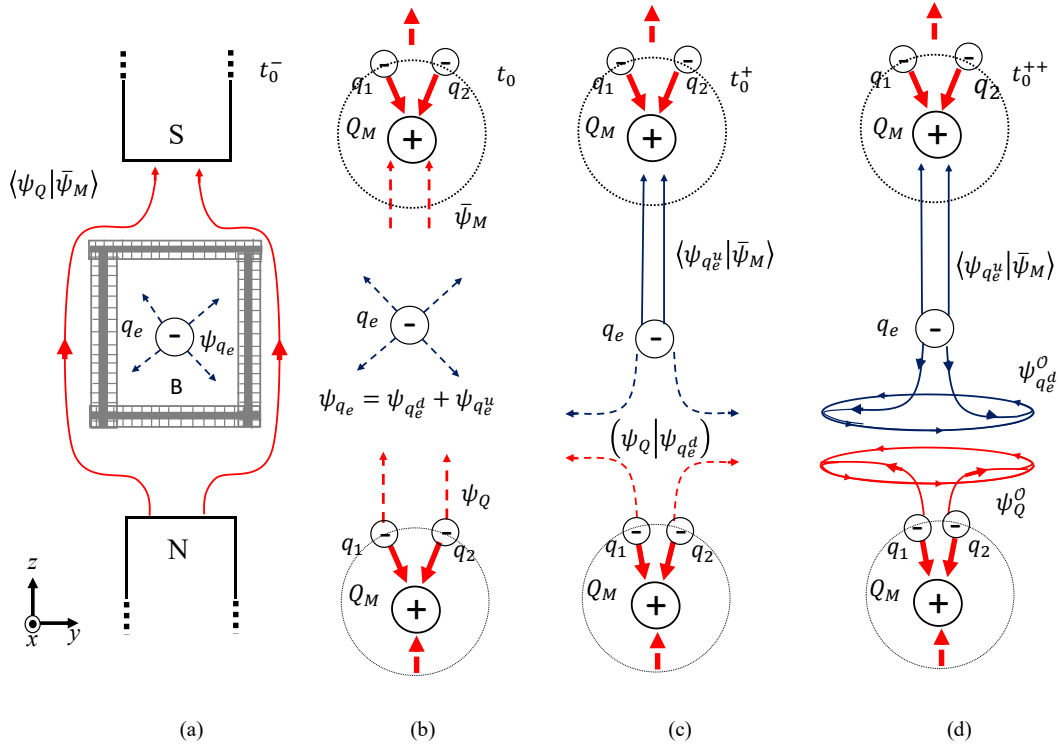

(a) (b) (c) (d)

Figure 3. (a) An electron with a total charge of $q_e$ and a corresponding string field function of $\psi_{q_e}$ enclosed by barrier B in macroscopic commutative SFI of the two poles of a dipole magnet (b) removal of barrier B at $t = t_0$ (c) and (d) noncommutative and commutative SFIs of the electron and the dipole magnet.

The two conjugated string field functions $\psi_Q$ and $\bar{\psi}_M$ in equation (A.3) are assigned to the string fields of the QC particles and the grand monopole of the dipole magnet (see Figure A.2 in the Appendix). With a total charge density of $\wp$ assigned to the north pole of the dipole magnet, a current of $\mathcal{J}^\mu$ is considered between the two conjugated string field functions, satisfying the conserved current associated with the $U(1)$ symmetry.

$$\partial_\mu \mathcal{J}^\mu = 0, \qquad \frac{\partial}{\partial t}\wp + \nabla \cdot \mathcal{J}^\mu(x^\mu) = 0 \qquad (3.1)$$

By taking the Lagrangian density invariant under $U(1)$ transformation, Noether's current can be calculated under an infinitesimal translation of the string fields $\psi(x^\mu) \to \psi(x^\mu) + \delta\psi(x^\mu)$, and subsequently, $\mathcal{L}(x^\mu) \to \mathcal{L}(x^\mu) + \delta\mathcal{L}(x^\mu)$[38].

$$\mathcal{L}(x^\mu) = \mathcal{L}\left(\psi(x^\mu), \partial_\mu \psi(x^\mu)\right) \qquad (3.2)$$

$$\delta\mathcal{L}(x^\mu) = \partial_\mu \left( \frac{\partial \mathcal{L}(x^\mu)}{\partial\left(\partial_\mu \psi(x^\mu)\right)} \delta\psi(x^\mu) \right) + \frac{\delta S}{\delta \psi(x^\mu)} \delta\psi(x^\mu) \qquad (3.3)$$

where the conserved current is calculated as follows:

$$\mathcal{J}^\mu(x^\mu) = \frac{\partial \mathcal{L}(x^\mu)}{\partial\left(\partial_\mu \psi(x^\mu)\right)} \delta\psi(x^\mu) \qquad (3.4)$$

The conserved current between the two conjugated string field functions can be derived by applying Noether's theorem, which satisfies the Klein-Gordon equation [38].

$$\mathcal{J}^\mu(x^\mu) = \bar{\psi}_M(x^\mu)\partial_\mu\psi_Q(x^\mu) - \psi_Q(x^\mu)\partial_\mu\bar{\psi}_M(x^\mu) \tag{3.5}$$

Similarly, the SFIs of an electron and the two conjugated Dirac string field functions associated with the two poles of the dipole magnet are analyzed to relate the spin of the electron string field to the microscopic disjoint and adjoint currents. The string field function of the electron, as a negatively charged particle, is represented as $\psi_{q_e}$, according to (2.1a).

For mathematical representation, the string field function of the electron is split into the string field functions $\psi_{q_e^d}$ and $\psi_{q_e^u}$ based on the quantum mechanical principle of linear superposition, in which the indices $d$ and $u$ indicate the half-down and half-up string fields of the function.

$$\psi_{q_e}(x^\mu) = \psi_{q_e^d}(x^\mu) + \psi_{q_e^u}(x^\mu) \tag{3.6}$$

As shown in Figure 3(b), the two poles of the dipole magnet are replaced by two permanently supersymmetry-broken G-groups of QC particles (see Section 4 and Figure A.2 in the Appendix). The two conjugated string field functions $\psi_Q$ and $\bar{\psi}_M$, corresponding to the QC particles and the grand monopole, interact with the string fields of the electron after the barrier B is removed at $t = t_0$. The string fields of the electron break the Hermitian symmetry of the commutative SFI of the two poles of the dipole magnet. The intrinsic background-independent phase difference of $\pi$ between the string fields of the electron and the grand monopole at the south pole and the phase difference of zero between the string fields of the electron and the QC particles at the north pole demonstrate commutative $\langle\psi_{q_e^u}|\bar{\psi}_M\rangle$ and noncommutative $(\psi_Q|\psi_{q_e^d})$ SFIs, respectively.

The commutative and noncommutative SFIs generate the corresponding adjoint $j^\mu(x)$ and disjoint $m_\mu(x)$ currents because of the zero and infinite resistances between the string fields of the two unlike- and like-charged quantum particles, respectively (see Figure 2(b) and (c)).

Inspection of Noether's theorem for the conserved current in (3.5) indicates that the conventional Lagrangian density includes only the adjoint current associated with the commutative SFI of the conjugated string field functions (3.7) [40]. However, as schematically shown in Figure 3(c), both adjoint and disjoint currents, correlated with positive and negative half-integer spins, are generated as a result of the commutative and noncommutative SFIs of the dipole magnet and the electron.

A comprehensive Lagrangian density should include both commutative and noncommutative SFIs corresponding to the adjoint and disjoint currents as the two parts of Majorana supercurrent [35]. The commutative and noncommutative SFIs are represented by conjugated and non-conjugated string field functions in which the first half-down and the second half-up quantized string fields of the electron, $\psi_{q_e^d}$ and $\psi_{q_e^u}$, are considered for the noncommutative and commutative SFIs with the string fields $\psi_Q$ and $\bar{\psi}_M$, respectively.

$$\mathcal{L}(x^\mu) = \mathcal{L}\left(\psi(x^\mu), \partial_\mu\psi(x^\mu)\right) \tag{3.7}$$

$$\mathcal{L}(x^\mu) = \mathcal{L}\left(\hat{\psi}_{q_e^u M}(x^\mu), \check{\psi}_{Q q_e^d}(x^\mu), \partial_\mu\hat{\psi}_{q_e^u M}(x^\mu), \partial^\mu\check{\psi}_{Q q_e^d}(x^\mu)\right) \tag{3.8}$$

where $\hat{\psi}_{q_e^u M} \equiv \langle\psi_{q_e^u}|\bar{\psi}_M\rangle$ and $\check{\psi}_{Q q_e^d} \equiv (\psi_Q|\psi_{q_e^d})$ are the commutative and noncommutative SFIs.

For an infinitesimal string field translation, the conjugated and non-conjugated string field functions associated with the commutative and noncommutative SFIs of the electron and the magnet can be represented as follows.

For the commutative SFIs,

$Ad_{Mq_e}: M \to q_e$

$$\psi_{q_e^u}(x^\mu) \to \psi_{q_e^u}(x^\mu) - i\alpha_{q_e^u M}\psi_{q_e^u}(x^\mu) \tag{3.9a}$$
$$\bar{\psi}_M(x^\mu) \to \bar{\psi}_M(x^\mu) + i\alpha_{Mq_e^u}\bar{\psi}_M(x^\mu) \tag{3.9b}$$

where $\alpha_{Mq_e^u} = \alpha_{q_e^u M} = \alpha_M$, according to the supersymmetry principle.

For the noncommutative SFIs,

$Di_{Qq_e}: Q \to q_e$

$$\psi_Q(x^\mu) \to \psi_Q(x^\mu) - i\beta_{Qq_e^d}\psi_Q(x^\mu) \quad (3.10a)$$

$$\psi_{q_e^d}(x^\mu) \to \psi_{q_e^d}(x^\mu) - i\beta_{q_e^d Q}\psi_{q_e^d}(x^\mu) \quad (3.10b)$$

where $\beta_{Qq_e^d} = \beta_{q_e^d Q} = \beta_Q$.

The coefficients $\alpha_M$ and $\beta_Q$ for the commutative and noncommutative SFIs are derived from the quantum number of the string fields in the interactions (see Appendix, Figure).

With the Lagrangian including both the commutative and noncommutative SFIs, the supercurrent is obtained as follows:

$$\mathcal{J}_{q_e}^\mu(x^\mu) = Im\left[\left(\psi_{q_e^u}(x^\mu)\overrightarrow{\partial_\mu}\bar{\psi}_M(x^\mu)\right) + \left(\psi_Q(x^\mu)\overleftrightarrow{\partial^\mu}\psi_{q_e^d}(x^\mu)\right)\right] \quad (3.11a)$$

or

$$\mathcal{J}_{q_e}^\mu(x^\mu) = \left[\alpha_M\left(\bar{\psi}_M(x^\mu)\partial_\mu\psi_{q_e^u}(x^\mu) - \psi_{q_e^u}(x^\mu)\partial_\mu\bar{\psi}_M(x^\mu)\right) \right. \\ \left. + \beta_Q\left(\psi_Q(x^\mu)\partial^\mu\psi_{q_e^d}(x^\mu) + \psi_{q_e^d}(x^\mu)\partial^\mu\psi_Q(x^\mu)\right)\right] \quad (3.11b)$$

For simplicity, the calculated current can be represented as follows:

$$\mathcal{J}_{q_e}^\mu(x^\mu) = \left[\partial_\mu\left(\alpha_M\hat{\psi}_{q_e^u M}(x^\mu)\right) + \partial^\mu\left(\beta_Q\check{\psi}_{q_e^d Q}(x^\mu)\right)\right] \quad (3.12)$$

where the $\hat{\psi}_{q_e^u M}$ and $\check{\psi}_{q_e^d Q}$ terms represent the commutative and noncommutative SFIs associated with the adjoint and disjoint currents, respectively.

The computed supercurrent in equation (3.12) includes the two terms of the partial differentials of both conjugated $\langle\psi_{q_e^u}|\bar{\psi}_M\rangle$ and non-conjugated $\left(\psi_Q|\psi_{q_e^d}\right)$ string field functions corresponding to the positive and negative half-integers of the electron spin. The conserved supercurrent associated with the charge density of the electron in the SFI, $\wp_e$ in (3.1), can be represented by the summation of the adjoint and disjoint currents.

$$\mathcal{J}_{q_e}^\mu(x^\mu) = j_e^\mu(x^\mu) + m_\mu^e(x^\mu) \quad (3.13)$$

where $j_e^\mu$ and $m_\mu^e$ are obtained as follows:

$$j_e^\mu(x^\mu) = Im\left(\psi_{q_e^u}(x^\mu)\overrightarrow{\partial_\mu}\bar{\psi}_M(x^\mu)\right) \quad (3.14a)$$

and

$$m_\mu^e(x^\mu) = Im\left(\psi_Q(x^\mu)\overleftrightarrow{\partial^\mu}\psi_{q_e^d}(x^\mu)\right) \quad (3.14b)$$

Based on the invariant Lagrangian under infinitesimal gauge string translation ($\delta\mathcal{L} = 0$) and equation (3.1), the microscopic adjoint and disjoint currents are dynamically exchanged.

$$\partial_\mu \mathcal{J}_{q_e}^\mu(x^\mu) = 0 \to \partial_\mu j_e^\mu(x^\mu) = -\partial^\mu m_\mu^e(x^\mu) \quad (3.15)$$

Figure 3(d) schematically shows the rotating vector string fields of the electron and the QC particles ($q_1$ and $q_2$) due to their noncommutative SFIs in the form of the closed string field functions $\psi_{q_e^d}^O$ and $\psi_Q^O$. The noncommutative SFIs contribute to a light cone gauge in non-abelian Yang-Mills theory [41], a holographic torus and other topological structures [11] and [42] (see Figure 14).

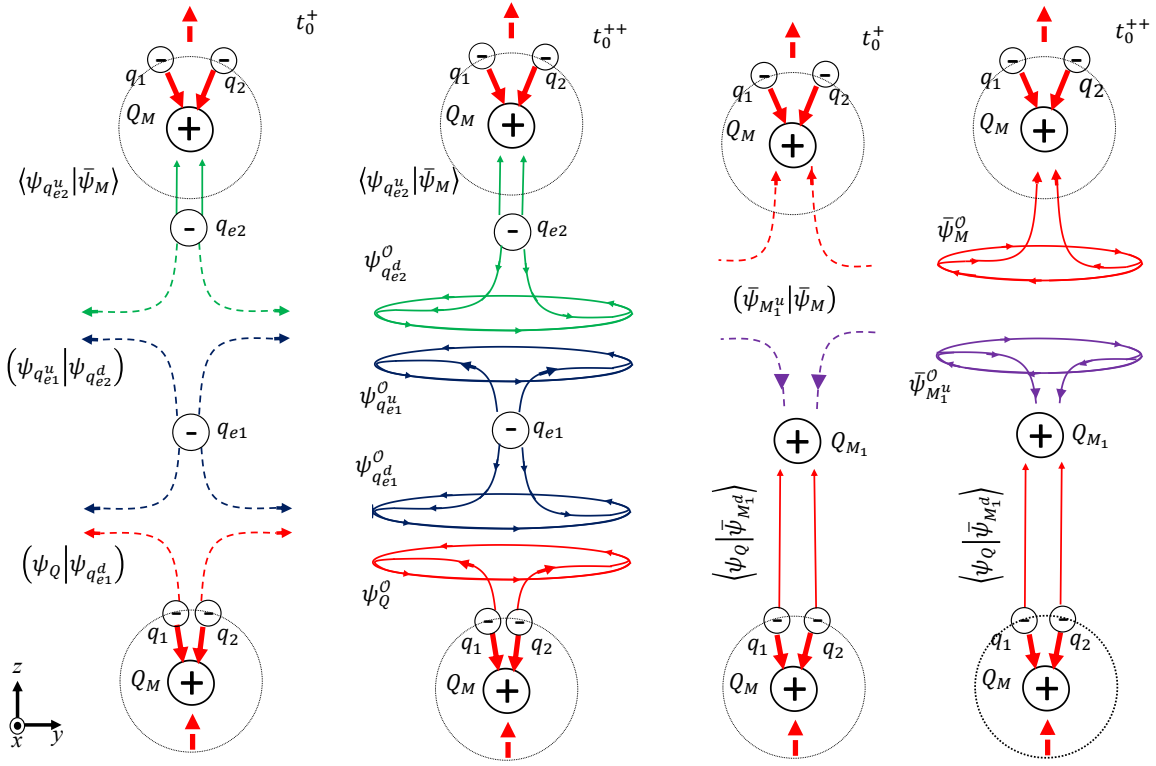

(a)                  (b)
Figure 4. (a) SFIs of two individual electrons and a dipole magnet (b) Spin of a grand monopole or a positively charged (PC) particle.

A question arises as to how the noncommutative string fields of the electron and the QC particles can demonstrate self-adjoint coupling. The issue requires a consideration of various physics principles, including supersymmetry, gravity, stability of the atomic structure, equibalance of the corresponding SFIs, and quantum vacuum. The manner in which rotating vector string fields arise in the form of topological tori is shown below (see Section 8). Other string fields of the two poles of the dipole magnet are not shown in Figure 3, but surround the noncommutative SFIs of the particles. These string fields contribute to the rotating vector string fields around the corresponding QC particles, including the electron.

The scenario can be generalized to SFIs with more than one electron and the conjugated field functions of a dipole magnet. Figure 4(a) schematically exemplifies the SFIs of two individual electrons and a dipole magnet.

Furthermore, the spin of a grand monopole or a positively charged (PC) particle can be verified. As schematically shown in Figure 4(b), a PC particle demonstrates the same commutative and noncommutative SFIs with the conjugated string field functions of the dipole magnet. The corresponding adjoint and disjoint currents can be similarly computed based on Noether's theorem.

## 4. Gauge group of QC particles; adjoint and disjoint currents

Thus far, the microscopic adjoint and disjoint currents have been discussed based on SFIs of unlike- and like - charged particles. The currents and corresponding commutative and noncommutative SFIs can be similarly characterized for a G-group of QC particles.

The characterization of a supersymmetric spatial geometry for a G-group of QC particles according to the 1/N expansion theory will further clarify the mutual relation between macroscopic electromagnetic fields and spacetime in the case of geometrical supersymmetry breaking.

Most importantly, a clear description of the unexplained magnetic current density in Maxwell's equations entails the compactification of both noncommutative and commutative SFIs in a G-group of QC particles in order to comply with Yang-Mills theory [9]. Maxwell himself was the first to recognize this necessity [43]. However, a G-group of QC particles as an atomic structure with supersymmetric spatial geometry is an underlying problem in modern physics [44]. This G-group should include not only QC (QED- and QCD-charged) particles but also their corresponding in-group commutative and noncommutative SFIs. Moreover, the G-group should include the corresponding intergroup commutative and noncommutative SFIs as well as gravitational interactions (see Section 8).

The 1/N expansion theory has been described as a promising approach for circumventing some mathematical complications in QCD, as suggested by 't Hooft [20]. Postulated by Edward Witten as an equivalency between quantum electrodynamics and quantum chromodynamics (electron-quark and photon-gluon) [44], the 1/N expansion is a potential solution for both circumventing mathematical difficulties and demonstrating a supersymmetric background-independent spatial geometry for a G-group of QC particles as an atomic structure. However, some issues remain for this approach.

Two vector string fields (lines of force between two quarks or electrons) cannot cross each other at the center because of their background-independent phase difference of zero; thus, the approach is limited to asymmetric diagrams for a large number of vertices ($N \to \infty$)[44]. Moreover, because the lines of force cannot cross each other, the diagrams demonstrate closed configurations, with no remaining space for the interior particles' lines of charge to interact with external fields (including the Higgs fields and intergroup SFIs). This issue prevents the compactification of both commutative and noncommutative SFIs in a single G-group of QC particles, which is necessary to demonstrate a comprehensive picture of abelian and non-abelian Yang-Mills theory [9].

This problem can be solved by taking the grand monopole as the identity element of a G-group of QC particles. By assigning one of the two conjugated fields (2.1b) to the grand monopole and the other field function (2.1a) to each QC particle in the G-group, both commutative and noncommutative SFIs can be incorporated in the G-group. The string fields of the grand monopole and each QC particle demonstrate commutative SFI, and the string fields of each pair of adjacent QC particles demonstrate noncommutative SFI in the G-group. If the grand monopole, as described by Polyakov and 't Hooft, is considered at the center, the supersymmetric planar and spherical diagrams would survive, in which all particles interact with the monopole without the crossing of string fields at the center.

The unification of the grand monopole with other fundamental interactions has already been studied in detail [6]. In this section, the noncommutative SFI between QED particles in the most exterior shells of a G-group are discussed and compactified in the model. The commutative and noncommutative SFIs are initially investigated for a two-particle semi-simple G-group of QC particles, with the PC particle as the identity element of the semi-simple G-group. The procedure is then generalized to an N-particle G-group of confined QC particles, with the grand monopole as the identity element of the G-group.

### 4.1. Semi-simple gauge group of QC particles

Figure 5(a) shows two NC particles and one PC particle with no initial SFI. Both commutative and noncommutative SFIs are considered for all isolated QC particles. Considering the width of $w_s$ for each string field [45], the string fields of the two NC particles can be argued to partially interact with the string fields of the PC particle because of the limited space around the PC particle and the repulsive interaction of the strings, according to Pauli's exclusion principle and divergent nature of string fields. This argument is more defensible if the 1/N expansion theory is considered for the SFIs of an infinite number of QC particles with the grand monopole at the center (see Figure 7).

Figure 5(b) schematically shows the SFIs of the QC particles. The string fields of the two quantum NC particles demonstrate a commutative SFI with the string fields of the PC particle, $\langle \psi_1 | \bar{\psi}_0 \rangle$ and $\langle \psi_2 | \bar{\psi}_0 \rangle$ (zero resistance between the strings and/or phase difference of $\pi$), and a noncommutative SFI with the string fields of each other, $(\psi_1 | \psi_2)$ (infinite resistance between the string and/or phase difference of zero).

It is postulated that the commutative and noncommutative SFIs generate Hermitian and anti-Hermitian operators, respectively. The commutative and noncommutative SFIs can be shown to have corresponding attractive, $\hat{\mathcal{F}}_{10}$ and $\hat{\mathcal{F}}_{20}$, and repulsive, $\tilde{\mathcal{F}}_{12}$, forces, and subsequently, each quantum NC particle exhibits two canonical momentum components in opposite directions [46]. By assigning the string field functions in (2.1a) and (2.1b) to the quantum NC and PC particles, respectively, the conjugate and nonconjugate relations between the string field functions of the unlike- and like-charged particles can be used to characterize the Hermitian and anti-Hermitian operators.

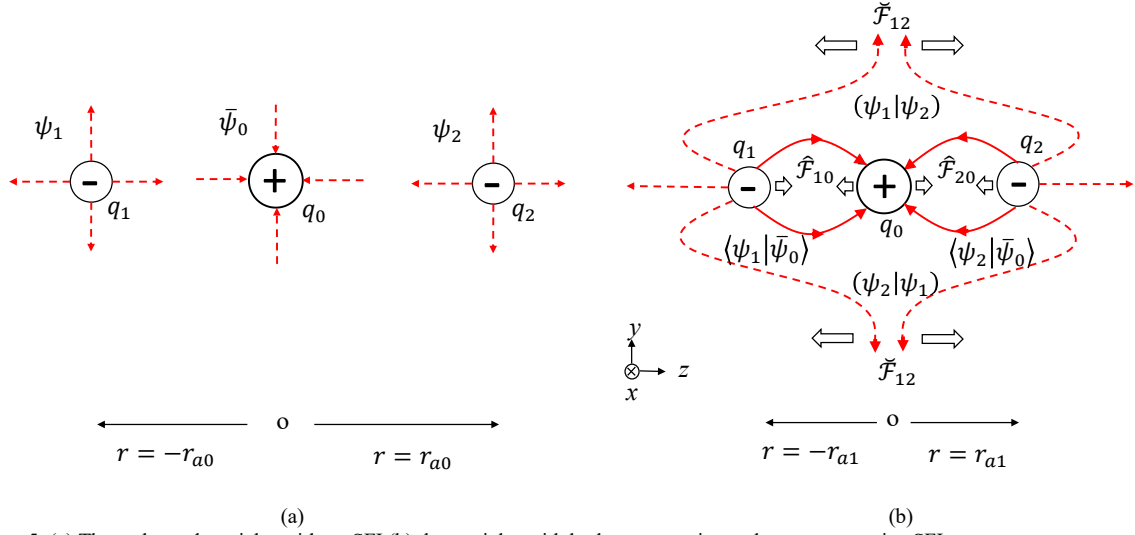

(a) (b)
Figure 5. (a) Three charged particles with no SFI (b) the particles with both commutative and noncommutative SFIs.

The nonrelativistic background-independent SFI between a pair of charged particles is generally considered as follows.

$$|\psi_{\pm q_{N1}}|A|\psi_{\pm q_{N2}}| = \{e^{\pm i\phi_{q_{N1}}}|\psi_{q_{N1}}|A|\psi_{q_{N2}}|e^{\pm i\phi_{q_{N2}}}\} \tag{4.1}$$

where the vertical bars represent the general type of SFI. Depending on the background-independent phase difference between string fields for a pair of charged particles, angled or curved brackets with separators are used for the commutative and noncommutative SFIs, respectively. Moreover, each string field function in (4.1) consists of background-independent and background-dependent phases; in this work, the latter, which correspond to relativistic field analysis, are not considered (see Section 8).

For the two QC particles with unlike charges of $q_1$ and $q_0$ shown in Figure 5, (4.1) represents the Hermitian conjugated operation.

$$\{e^{-i\phi_{q_1}}|\psi_{q_1}|A|\psi_{q_0}|e^{+i\phi_{q_0}}\} \equiv \langle\psi_{q_1}|\hat{A}|\bar{\psi}_{q_0}\rangle \tag{4.2}$$

In contrast, for the two particles with like charges of $q_1$ and $q_2$ shown in Figure 5, (4.1) represents the anti-Hermitian nonconjugated operation.

$$\{e^{-i\phi_{q_1}}|\psi_{q_1}|A|\psi_{q_2}|e^{-i\phi_{q_2}}\} \equiv (\psi_{q_1}|\breve{A}|\psi_{q_2}) \tag{4.3}$$

The different background-independent phase differences between the string fields of a pair of unlike- or like-charged particles is shown to determine the Hermitian or anti-Hermitian operations of the particles SFIs.

$$\overline{\{e^{-i\phi_{q_{N1}}}|\psi_{q_{N1}}|A|\psi_{q_{N2}}|e^{+i\phi_{q_{N2}}}\}} = \{e^{+i\phi_{q_{N1}}}|\psi_{q_{N1}}|A|\psi_{q_{N2}}|e^{-i\phi_{q_{N2}}}\} \equiv \langle\psi_{q_{N1}}|\hat{A}|\bar{\psi}_{q_{N2}}\rangle \tag{4.4a}$$

$$\begin{aligned}\overline{\{e^{-i\phi_{q_{N1}}}|\psi_{q_{N1}}|A|\psi_{q_{N2}}|e^{-i\phi_{q_{N2}}}\}} &= \{e^{+i\phi_{q_{N1}}}|\psi_{q_{N1}}|A|\psi_{q_{N2}}|e^{+i\phi_{q_{N2}}}\} \\ &= -\{e^{-i\phi_{q_{N1}}}|\psi_{q_{N1}}|A|\psi_{q_{N2}}|e^{-i\phi_{q_{N2}}}\} \equiv (\psi_{q_{N1}}|\breve{A}|\psi_{q_{N2}})\end{aligned} \tag{4.4b}$$

where the hat and breve operators represent Hermitian (adjoint) and anti-Hermitian (disjoint) operations. The adjoint and disjoint currents corresponding to the commutative and noncommutative SFIs between the particles can then be calculated (see Section 3).

### 4.2. Gauge group of infinite numbers of QC particles

The semi-small G-group of two quantum NC particles interacting with one PC particle in Figure 5 can be extended to larger numbers of NC particles, e.g., for an arbitrary number $N$ of negatively charged particles, with $N$ approaching infinity.

According to the 1/N expansion theory, an infinite number of QED particles can be considered in the most exterior shell of a G-group. Figure 6(a) shows the grand monopole and four NC particles, where the number of NC particles can be increased to infinity ($N \to \infty$). The key aspect here is that the spatial geometry of the G-group is the fundamental constraint on the number of string fields of each NC particle in the commutative SFI with the string fields of the grand monopole at the core. This spatial limitation causes the string fields of each particle to partially interact with the string fields of the monopole, other adjacent QC particles in the G-group, and NC and PC particles in other adjacent G-groups (see Section 8).

The following discussion assumes that there is no initial SFI between the string field functions of the four NC particles, $\psi_n$, ($n = 1,2,3,4$), and the grand monopole, $\bar{\psi}_M$. Following common conventions, the four NC particles with identical negative charges are initially as far as possible from one another in their normal positions.

Each NC particle demonstrates both a commutative SFI with the grand monopole (identity element) and a noncommutative SFI with the adjacent NC particles. For instance, the NC particle with a charge of $q_1$ is subject to commutative, $\hat{\psi}_{1M} \equiv \langle \psi_1 | \bar{\psi}_M \rangle$, and noncommutative, $\check{\psi}_{12} \equiv (\psi_1 | \psi_2)$ and $\check{\psi}_{41} \equiv (\psi_4 | \psi_1)$, SFIs with the grand monopole and the two adjacent NC particles with charges of $q_2$ and $q_4$, respectively.

The NC particles are pulled by the grand monopole toward the center due to their commutative SFI; however, each NC particle is pushed by other adjacent NC particles due to their noncommutative SFI.

According to Coulomb's law, the more the NC particles approach the grand monopole at the center, the distance between each pair of adjacent NC particles decreases, resulting in a greater repulsive force between the NC particles.

Overall, each NC particle with string field function $\psi_{q_N}$ experiences attractive and repulsive canonical momentum components in opposite directions because of the translations of the commutative and noncommutative SFIs corresponding to the grand monopole, $\bar{\psi}_M$, and other adjacent NC particles, $\psi_{q_{N-1}}$ and $\psi_{q_{N+1}}$.

$$\langle \bar{\psi}_M(x^\mu) | \hat{p} | \psi_N(x^\mu) \rangle = i\hbar \partial_\mu \hat{\psi}_{MN}(x^\mu) \quad (4.5a)$$
$$(\psi_{N-1,N+1}(x^\mu) | \check{p} | \psi_N(x^\mu)) = -i\hbar \left( \partial^\mu \check{\psi}_{N-1,N}(x^\mu) + \partial^\mu \check{\psi}_{N+1,N}(x^\mu) \right) \quad (4.5b)$$

The canonical momentum components are assumed to act inward and outward, with the G-group as the frame of reference. Assuming equal SFIs between each NC particle and the grand monopole, the probable supersymmetric spatial geometry for the G-group is a circle in two dimensions and/or a sphere in three dimensions, with a stable formation of NC particles around the grand monopole in the G-group; e.g., according to Newton's third law, the outward repulsive force $\vec{\mathcal{F}}_2 = \vec{\mathcal{F}}_{12} + \vec{\mathcal{F}}_{32}$ on particle $N = 2$ resulting from the canonical repulsive momentum in (4.5b) is counterbalanced by the inward attractive force $\vec{\mathcal{F}}_{2M}$ resulting from the canonical attractive momentum in (4.5a) [Figure 6(b)].

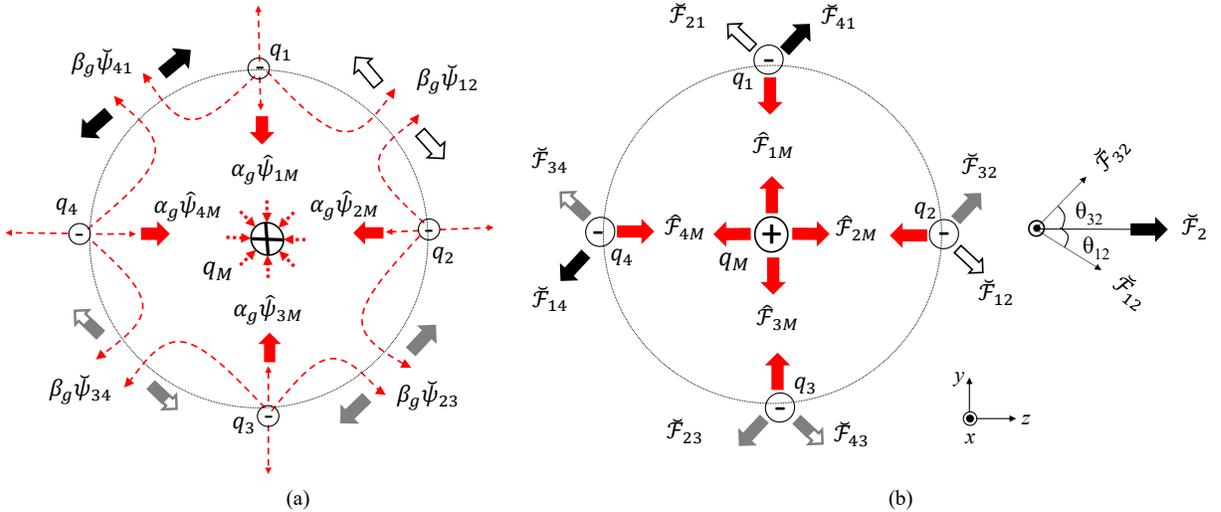

(a)          (b)
Figure 6. (a) Commutative SFI of N negatively charged particles with a grand monopole accompanied by their noncommutative SFI (b) Inward-attractive and outward-repulsive forces on the QC particles.

Figure 7(a) shows the 2D supersymmetric spatial geometry of the G-group according to the commutative SFI between the NC particles and the grand monopole as well as the noncommutative SFI between the NC particles [2].

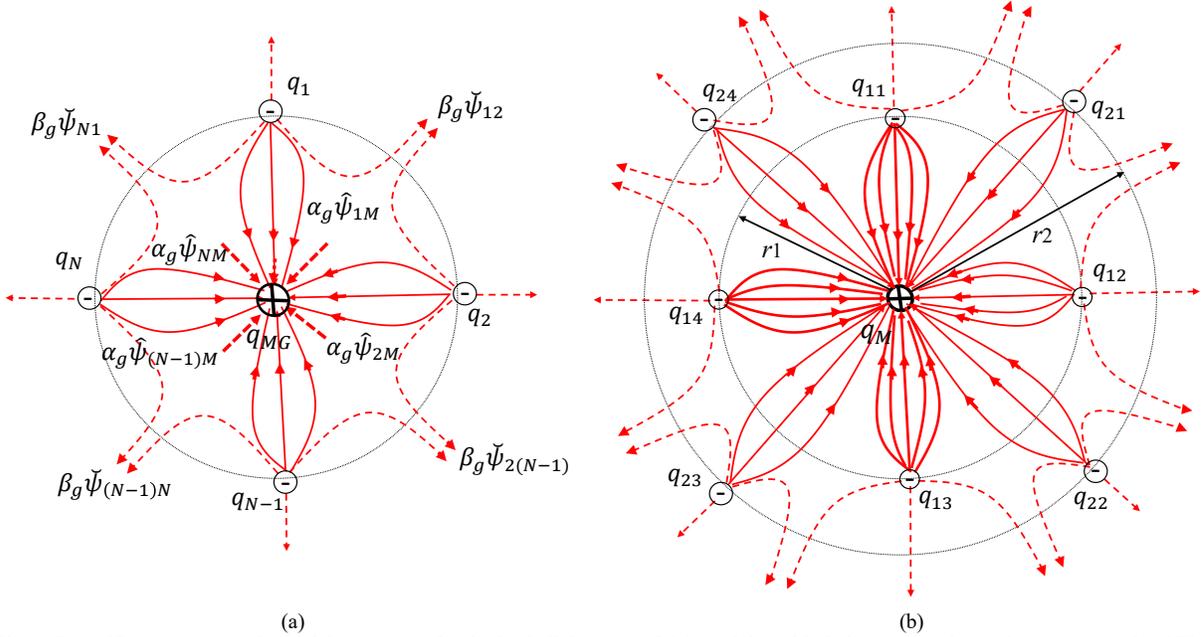

(a)                                                                                           (b)

Figure 7. (a) 2D supersymmetric spatial geometry of a single-shell G-group of QC particles with their commutative and noncommutative SFIs (b) 2D supersymmetric spatial geometry of a dual-shell G-group of QC particles.

The adjoint and disjoint currents corresponding to the commutative and noncommutative SFIs in the G-group are similarly determined by assigning a pair of conjugated field functions to the string fields of each NC particle and the grand monopole and a pair of nonconjugated field functions to the string fields of a pair of adjacent QC particles in the G-group. Equations (4.6a) and (4.6b) represent the infinitesimal quantum number of commutative SFIs between the $n$th NC particle and the grand monopole.

$Ad_G: G \to G$

$$\psi_n(x^\mu) \to \psi_n(x^\mu) - i\alpha_{ng}\psi_n(x^\mu) \quad (4.6a)$$
$$\bar{\psi}_M(x^\mu) \to \bar{\psi}_M(x^\mu) + i\alpha_{nM}\bar{\psi}_M(x^\mu) \quad (4.6b)$$

where $\alpha_{ng} = \alpha_{nM}$, proportional to the quantum number of string fields of the $n$th NC particle in the commutative SFI with the grand monopole (see Section 8 and Appendix, Figure). The infinitesimal quantum number of gauge commutative SFIs between all the NC particles and the grand monopole in the G-group is represented as follows.

$$\sum_{n=1}^{N}\psi_n(x^\mu) \to \sum_{n=1}^{N}\psi_n(x^\mu) - \alpha_G\sum_{n=1}^{N}\psi_n(x^\mu) \quad (4.6c)$$

where $\alpha_G = \sum_{n=1}^{N}\alpha_{ng}$ and $\alpha_M = \sum_{n=1}^{N}\alpha_{nM}$. Moreover, $\alpha_{ng} = \alpha_g$ according to the supersymmetry principle.

$Di_G: G \to G$

For a pair of adjacent NC particles in the G-group with an infinitesimal quantum number of noncommutative SFI:

$$\psi_n(x^\mu) \to \psi_n(x^\mu) - i\beta_{ng}\psi_n(x^\mu) \quad (4.7a)$$
$$\psi_{n+1}(x^\mu) \to \psi_{n+1}(x^\mu) - i\beta_{(n+1)g}\psi_{n+1}(x^\mu) \quad (4.7b)$$

where $\beta_{ng} = \beta_{(n+1)g} = \beta_g$, based on the supersymmetry condition of the G-group spatial geometry.

The Lagrangian density is compactly constructed for the $N$-NC-particle G-group in terms of the commutative and noncommutative SFIs.

$$\mathcal{L}_G(x^\mu) = \bar{\psi}_M(x^\mu)\overrightarrow{\partial_\mu}\left(\sum_{n=1}^{N}\psi_n(x^\mu)\right) - \left(\psi_1(x^\mu)\overleftrightarrow{\partial^\mu}\psi_N + \sum_{n=1}^{N-1}\left(\psi_n(x^\mu)\overrightarrow{\partial^\mu}\psi_{n+1}(x^\mu)\right)\right) \tag{4.8}$$

where the first term and the other two terms represent the commutative and noncommutative SFIs corresponding to the total microscopic adjoint and disjoint currents of the G-group, respectively. The supercurrent of the G-group is derived by applying Noether's theorem (3.3).

$$\begin{aligned}\mathcal{J}_G^\mu(x^\mu) &= j_G^\mu(x^\mu) + m_\mu^G(x^\mu) \\ &= \alpha_G\left[\left(\sum_{n=1}^{N}\psi_n(x^\mu)\right)\partial_\mu\bar{\psi}_M(x^\mu) - \bar{\psi}_M(x^\mu)\partial_\mu\left(\sum_{n=1}^{N}\psi_n(x^\mu)\right)\right] \\ &+ \beta_G\left[\sum_{n=1}^{N}\left(\psi_n(x^\mu)\partial^\mu\psi_{n+1}(x^\mu) + \psi_{n+1}(x^\mu)\partial^\mu\psi_n(x^\mu)\right)\right]\end{aligned} \tag{4.9}$$

The supercurrent of the G-group including the adjoint and disjoint currents can be represented as follows:

$$\mathcal{J}_G^\mu(x^\mu) = j_G^\mu(x^\mu) + m_\mu^G(x^\mu) = \sum_{n=1}^{N} j_{nM}^\mu(x^\mu) + \sum_{n=1}^{N} m_\mu^{(n,n+1)}(x^\mu) \tag{4.10}$$

The invariant Lagrangian condition under infinitesimal translation implies the conservation of current in which the adjoint and disjoint currents are dynamically exchanged as a consequence of supersymmetry breaking (diffeomorphism) of the G-group (see Section 5).

$$\partial_\mu \mathcal{J}_G^\mu(x^\mu) = 0 \rightarrow \partial_\mu\left(\sum_{n=1}^{N} j_{nM}^\mu(x^\mu)\right) = -\partial^\mu\left(\sum_{n=1}^{N} m_\mu^{(n,n+1)}(x^\mu)\right) \tag{4.11}$$

Moreover, the G-group of NC particles can be developed as a multi-shell structure, as illustrated in Figure 7(b). The supercurrent can be similarly determined for a multi-shell G-group of NC particles, in which the NC particles in the exterior and interior shells are associated with lower and higher current densities in the G-group (QED and QCD particles). Furthermore, the 3D spatial geometry of the G-group structure can be developed in a similar manner.

## 5. Spontaneous supersymmetry breaking in the spatial geometry of a gauge group of QC particles

In this section, spontaneous supersymmetry breaking in the spatial geometry of the G-group of QC particles is discussed. This phenomenon is characterized by the SFIs of an external supercharge with constant zero-mode string fields and a G-group (constant B-field in [10]). Figure 8 shows two QC particles with a string field function of $\psi_Q$, a charge density of $\wp_Q$, and a corresponding current of $\mathcal{J}_Q^\mu$ under the Q-subgroup nomenclature in the vicinity of a G-group with no initial SFIs. A barrier B is assumed between the Q-subgroup and the G-group, indicating the absence of SFIs. The string fields of the Q-subgroup are generally the string fields of $N_\gamma$ QC particles at one pole of a dipole magnet (see Figure A.2 in the Appendix). After the barrier B is removed, the Q-subgroup and G-group demonstrate both noncommutative (anti-Hermitian) and commutative (Hermitian) SFIs.

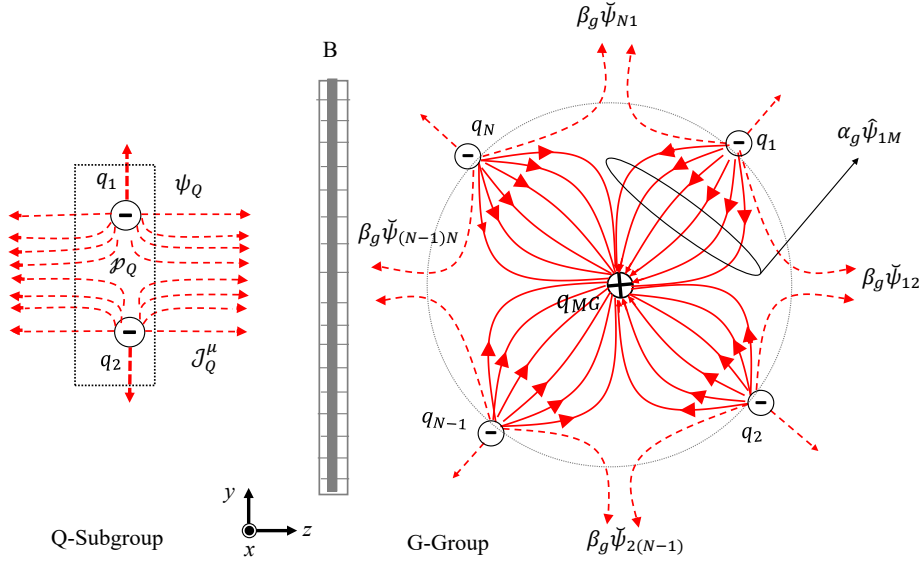

Figure 8. Q-subgroup and a 2D supersymmetric G-group of QC particles.

The right-hand string field function of the Q-subgroup is split into two string field functions, $\psi_{Q_A}$ and $\psi_{Q_D}$, based on the quantum mechanical principle of linear superposition, in which the commutative and noncommutative SFIs and their corresponding adjoint and disjoint currents are determined.

$$\psi_Q(x^\mu) = \psi_{Q^A}(x^\mu) + \psi_{Q^D}(x^\mu), \psi_{Q^A}(x^\mu) = \alpha_Q \psi_Q(x^\mu), \psi_{Q^D}(x^\mu) = \beta_Q \psi_Q(x^\mu) \tag{5.1}$$

where the coefficients $\alpha_Q$ and $\beta_Q$ associated with the commutative and noncommutative SFIs of the Q-subgroup and G-group are determined from the quantum number of the string fields in the interactions (see Appendix, Figure). The commutative and noncommutative SFIs are represented as the actions of linear string field functions.

$$\langle(\psi_Q(x^\mu)||\psi_G(x^\mu))\rangle = \langle \psi_{Q^A}(x^\mu)|\bar{\psi}_{MG}(x^\mu)\rangle + \left(\psi_{Q^D}(x^\mu)\Big|\check{\psi}_{N(N-1)}(x^\mu)\right) \tag{5.2}$$

where the notation $\langle(||)\rangle$ is used to represent both commutative and noncommutative SFIs. Moreover, as schematically shown in Figure 8, $\bar{\psi}_{MG}$ and $\check{\psi}_{N(N-1)}$ represent the string field functions of the grand monopole and the noncommutative SFIs of two adjacent QC particles in the G-group, respectively.

In a demonstration of the commutative SFI with the grand monopole of the G-group, the Q-subgroup's string fields ($\psi_Q$) encounter resistance from the QC particles of the G-group. The resistance is caused by the noncommutative SFI of the QC particles of the Q-subgroup and G-group due to the constant background-independent phase of the QC particles string fields in the Q-subgroup and G-group. [the exterior noncommutative SFI of the QC $q_{N-1}$ and $q_N$ particles ($\beta_G \check{\psi}_{N(N-1)}$] (see Figure 2, Figure 8, and Figure 9). In the 2D spatial geometry of the G-group, two QC particles are considered for the noncommutative SFI with the external Q-subgroup's string fields; however, the scenario would differ for the 3D spatial geometry of the G-group of QC particles.

As shown in Figure 9(a), some of the string fields of the Q-subgroup initially interact with the noncommutative string fields of QC particles in the G-group. The $\psi_{Q^D}$ string field function acts on the string fields of two QC particles in the G-group ($q_{N-1}$ and $q_N$), where the two actions can be separately constructed according to the quantum mechanical principle of linear superposition.

$$\left(\psi_{Q^D}(x^\mu)\Big|\check{\psi}_{N(N-1)}(x^\mu)\right) = \left(\frac{1}{2}\psi_{Q^D}(x^\mu)\Big|\psi_{N-1}(x^\mu)\right) + \left(\frac{1}{2}\psi_{Q^D}(x^\mu)\Big|\psi_N(x^\mu)\right) \tag{5.3}$$

Due to the noncommutative SFI, the string fields of both the Q-subgroup and G-group in the SFI experience additional tension, which can be determined from the equations derived for the strings of the particles (see Figure 2(b)) [47].

$$\mathcal{L} = \int \left( \frac{1}{2} \mu_0 dx \left(\frac{dy}{dx}\right)^2 - \frac{1}{2} T_0 dx \left(\frac{dy}{dx}\right)^2 \right) \tag{5.4}$$

where $\mu_0$ is the mass of the quantum scale string per unit length.

The string fields of each QC particle in the G-group are under a constant tension $T_0$ due to the in-group noncommutative SFI between two adjacent QC particles [e.g., QC particles with charges of $q_{N-1}$ and $q_N$ ($\beta_g \check{\psi}_{N(N-1)}$)]. As geometrically illustrated in Figure 9, the initial constant tensions on the string fields associated with the two QC particles increase due to the noncommutative SFI of the Q-subgroup and G-group (5.10).

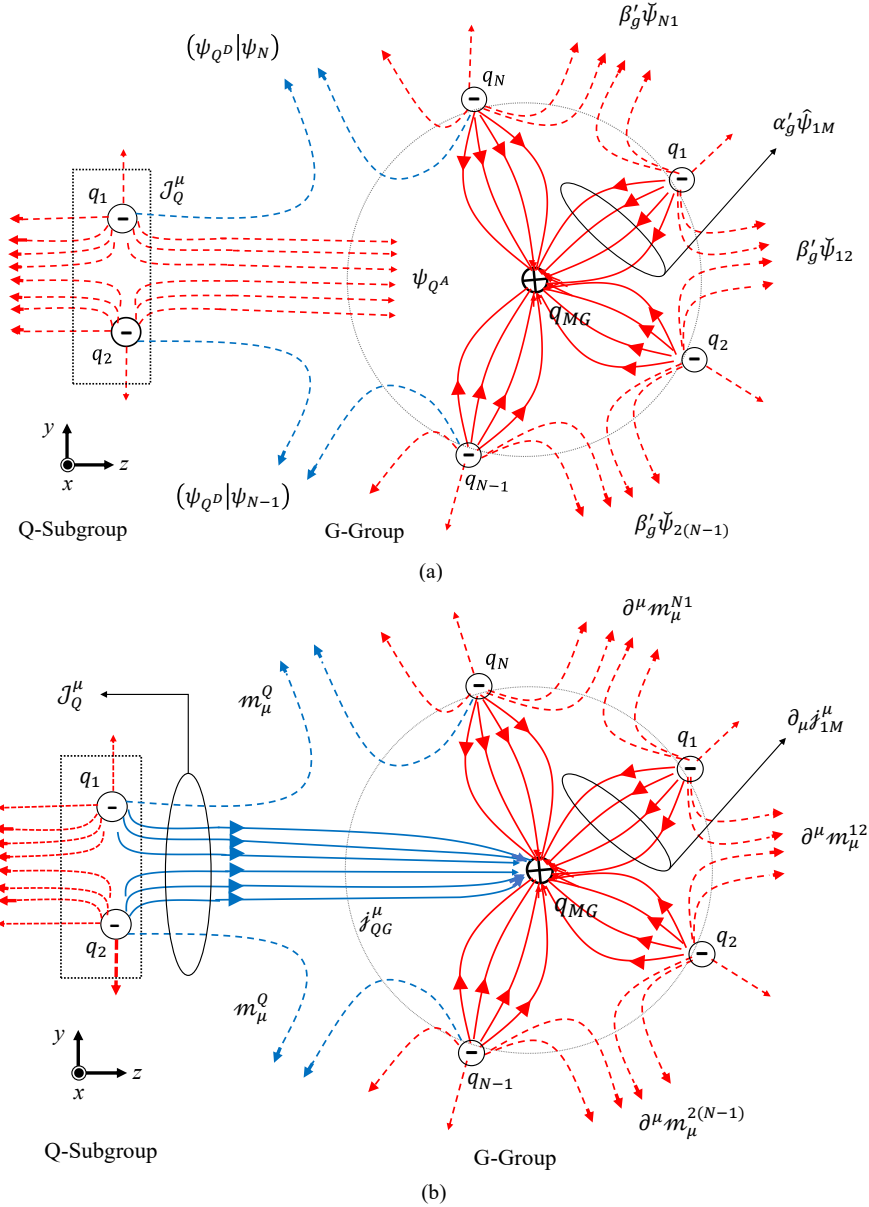

Figure 9. Supercharge of Q and a G-group of QC particles with (a) noncommutative SFI and (b) noncommutative and commutative SFIs.

The change in tension on the two QC particles in the G-group generates canonical anti-Hermitian (repulsive) momentum components acting on the two QC particles.

Consideration of the confinement of QC particles with the grand monopole at the center indicates that the canonical anti-Hermitian momentum components applied to the two QC particles cause the particles to rotate around the core ($\ell \leq \frac{\hbar}{mc}$ [48]). As schematically shown in Figure 9, the $q_{N-1}$ and $q_N$ QC particles are squeezed due to the imposed canonical anti-Hermitian momenta of the noncommutative SFIs $\beta'_m \check{\psi}_{QG}$. Consequently, the constant tension between the two QC particles and their nearest-neighbor particles increases ($\beta_g \check{\psi}_{N1} \to \beta'_g \check{\psi}_{N1}$, $\beta_g \check{\psi}_{2(N-1)} \to \beta'_g \check{\psi}_{2(N-1)}$ and $\beta_g \check{\psi}_{12} \to \beta'_g \check{\psi}_{12}$).

Figure 9(a) shows that the supersymmetry of the G-group spatial geometry is spontaneously broken due to the noncommutative SFI of the QC particles in the Q-subgroup and G-group.

As schematically shown in Figure 9(b), a commutative SFI arises between the string fields of the QC particles of the Q-subgroup and the grand monopole of the G-group. The adjoint and disjoint currents are computed for infinitesimal quantum numbers of commutative and noncommutative SFIs between the Q-subgroup and G-group, thus leaving the Lagrangian invariant.

### 5.1. Commutative and noncommutative SFIs of Q-subgroup and G-group, adjoint and disjoint currents

A pair of conjugated field functions $\langle \psi_{Q^A} | \bar{\psi}_M \rangle$ and a pair of nonconjugated field functions $(\psi_{Q^D} | \psi_n)$ are assigned to the commutative and noncommutative SFIs to determine the corresponding microscopic adjoint and disjoint currents, respectively, between the Q-subgroup and G-group. For an infinitesimal gauge quantum string field translation from the Q-subgroup to the G-group, the conjugated and nonconjugated field functions represent the commutative and noncommutative SFIs.

$Ad_{QG}: Q \to G$

$$\psi_{Q^A}(x^\mu) \to \psi_{Q^A}(x^\mu) - i\gamma_Q^A \psi_{Q^A}(x^\mu) \quad (5.5a)$$
$$\bar{\psi}_M(x^\mu) \to \bar{\psi}_M(x^\mu) + i\gamma_G^A \bar{\psi}_M(x^\mu) \quad (5.5b)$$

where $\bar{\psi}_M$ is the field function of the string fields of the grand monopole in the G-group, which commutatively interacts with the string fields of the QC particles in the Q-subgroup, and $\gamma_Q^A = \gamma_G^A$ are the coefficients of the Q-subgroup and G-group, respectively, which are proportional to the quantum number of their string fields in the intergroup commutative (adjoint) SFI (see Section 8 and Figure A.1 in the Appendix).

$Di_{QG}: Q \to G$

$$\psi_{Q^D}(x^\mu) \to \psi_{Q^D}(x^\mu) - i\gamma_Q^D \psi_{Q^D}(x^\mu) \quad (5.6a)$$

$$\sum_{n=1}^{N_\gamma^D} \psi_n(x^\mu) \to \sum_{n=1}^{N_\gamma^D} \psi_n(x^\mu) - i\gamma_G^D \sum_{n=1}^{N_\gamma^D} \psi_n(x^\mu) \quad (5.6b)$$

where $\sum_{n=1}^{N_\gamma^D} \psi_n(x^\mu)$ is the string field function of the QC particles in the G-group that noncommutatively interact with the QC particles in the Q-subgroup and $\gamma_Q^D = \gamma_G^D$ are the coefficients of the Q-subgroup and G-group, respectively, which are proportional to the quantum number of the string fields in the intergroup noncommutative (disjoint) SFIs (see Section 8).

Moreover, $N_\gamma^D$ is the number of QC particles in the G-group whose string fields partially participate in the noncommutative SFIs. Notably, the number of QC particles in the G-group is assumed to be equal to two, $N_\gamma^D = 2$, for a 2D single-shell G-group spatial geometry; however, a large number of QC particles in the G-group partially participate in the interaction for a 3D multi-shell G-group spatial geometry.

The Lagrangian density can be compactly constructed, including both the commutative and noncommutative SFIs.

$$\mathcal{L}_{QG}(x^\mu) = \psi_{Q^A}(x^\mu) \overleftrightarrow{\partial_\mu} \bar{\psi}_M(x^\mu) - \psi_{Q^D}(x^\mu) \overleftrightarrow{\partial^\mu} \left( \sum_{n=1}^{N_\gamma^D} \psi_n(x^\mu) \right) \quad (5.7)$$

By applying Noether's theorem (3.3), the conserved supercurrent translated from the Q-subgroup to the G-group is derived.

$$\mathcal{J}_Q^\mu(x^\mu) = j_{QG}^\mu(x^\mu) + m_\mu^{QG}(x^\mu)$$

$$= \left[ \gamma_Q^A \left( \bar{\psi}_M(x^\mu) \partial_\mu \psi_{Q^A}(x^\mu) - \psi_{Q^A}(x^\mu) \partial_\mu \bar{\psi}_M(x^\mu) \right) \right.$$

$$\left. + \gamma_Q^D \left( \psi_{Q^D}(x^\mu) \partial^\mu \left( \sum_{n=1}^{N_\gamma^D} \psi_n(x^\mu) \right) + \left( \sum_{n=1}^{N_\gamma^D} \psi_n(x^\mu) \right) \partial^\mu \psi_{Q^D}(x^\mu) \right) \right] \quad (5.8)$$

The supercurrent translated from the Q-subgroup to the G-group includes both the adjoint $j_{QG}^\mu = Im(\psi_{Q^A} \overleftrightarrow{\partial_\mu} \bar{\psi}_M)$ and disjoint $m_\mu^{QG} = Im\left(\psi_{Q^D} \overleftrightarrow{\partial^\mu} \sum_{n=1}^{N_\gamma^D} \psi_n\right)$ currents. The condition of an invariant Lagrangian under infinitesimal translation implies conservation of the supercurrent.

$$\partial_\mu \mathcal{J}_Q^\mu(x^\mu) = 0 \rightarrow \partial_\mu j_{QG}^\mu(x^\mu) = -\partial^\mu m_\mu^{QG}(x^\mu) \quad (5.9)$$

Due to the infinitesimal gauge translation from the Q-subgroup to the G-group, the geometrical supersymmetry of the G-group is broken, as shown in Figure 9. The geometrical supersymmetry breaking in the G-group weakens the commutative string fields of the G-group QC particles coupled to the grand monopole (5.10), which is associated with the strong in-group SFI of the G-group (see Section 8). This phenomenon can be associated with the asymptotic freedom principle in the QCD theory of sub-atomic particles [17].

This phenomenon is schematically illustrated in Figure 9. Comparison of the symmetric and asymmetric spatial geometries of the G-group in Figure 8 and Figure 9 indicates that the quantum number of the string fields of each QC particle in the commutative SFI with the grand monopole decreases while the quantum number of the corresponding string fields in the noncommutative SFI with the adjacent QC particles in the G-group increases.

$$\alpha_g \hat{\psi}_{nM}(x^\mu) \rightarrow \alpha_g' \hat{\psi}_{nM}(x^\mu) \quad \alpha_g > \alpha_g' \quad (5.10a)$$
$$\beta_g \check{\psi}_{n-1,n}(x^\mu) \rightarrow \beta_g' \check{\psi}_{n-1,n}(x^\mu) \quad \beta_g < \beta_g' \quad (5.10b)$$

The mechanism by which the geometrical supersymmetry breaking of the G-group alters the amplitude (the number of strings) of the commutative SFI between the QC particles and the grand monopole as well as that of the noncommutative SFI between each pair of QC particles in the G-group was evaluated for corresponding adjoint and disjoint currents. The derived equations for the conserved current confirm the application of the asymptotic freedom principle to the SFIs of the G-group, through which the adjoint and disjoint currents corresponding to the commutative and noncommutative SFIs are dynamically exchanged. As the translation of gauge fields increases from the Q-subgroup to the G-group, the amplitude of the commutative SFI between the QC particles and the grand monopole decreases, with a subsequent increase in amplitude of the noncommutative SFI between each pair of adjacent QC particles in the G-group.

In contrast, a quantum ratio of the disjoint current of the G-group is translated to the next G-group ($\mathcal{J}_{GG}^\mu$), and the remainder of the current demonstrates a self-adjoint coupled string field ($m_G^O$) in the shape of a torus with compactified commutative and noncommutative SFIs (see Figure 10 and Figure 14). The disjoint current $m_\mu^G$ should also be conserved and invariant under the gauge string field translation.

$$m_\mu^G(x^\mu) = m_G^O(x^\mu) + \mathcal{J}_{GG}^\mu(x^\mu) \quad (5.11a)$$
$$\partial^\mu m_\mu^G(x^\mu) = 0 \rightarrow \partial^\mu m_G^O(x^\mu) = -\partial_\mu \mathcal{J}_{GG}^\mu(x^\mu) \quad (5.11b)$$

# 6. Magnetic current density in Maxwell's equations

The adjoint microscopic current $j^\mu$ is associated with the commutative SFIs $\langle\psi|\bar\psi\rangle$ in quantum mechanics [38]. In electromagnetic field interactions, the same current is associated with a macroscopic electric field, according to the vector form of Ohm's law $j = \sigma E$ [49]. In contrast, the disjoint microscopic current $m_\mu$ (discussed in Section 5) is attributed to the noncommutative SFI of two adjacent QC particles in a G-group of QC particles.

As discussed in Section 2 for the microscopic SFIs of unlike- and like-charged QC particles (Figure 2), no closed string of charge is confined between two like-charged quantum particles, implying an infinite resistance between the particles (Dirichlet-Nauman boundary conditions for the strings of charges). However, closed strings of charge are confined between two unlike-charged particles, implying an absolutely zero resistance between the particles (Dirichlet-Dirichlet boundary conditions for the strings of charges). Therefore, the string field analysis of quantum particles provides two discrete absolutely zero and infinite resistances.

By considering the Q-subgroup as a subgroup of the G-group (see Figure), the two absolutely zero and infinite resistances corresponding to microscopic commutative and noncommutative SFIs of QC particles are shown to take finite values, greater than zero and less than infinity, for the macroscopic intergroup commutative and noncommutative SFIs.

As discussed in Section 5, the adjoint and disjoint currents $j^\mu_{QG}$ and $m_\mu^{QG}$ are correlated with the resistivity of the G-group. The string fields associated with the adjoint current $j^\mu_{QG}$ are confined as closed string fields between the QC particles of a Q-subgroup and the grand monopole of the G-group, and those associated with the disjoint current $m_\mu^{QG}$ demonstrate circular closed string fields around the G-group.

Section 5 demonstrated that the SFIs of the Q-supercharge and the G-group result in spontaneous supersymmetry breaking of the G-group spatial geometry. Subsequently, the infinitesimal movement of the QC particles ($\ell \leq \frac{\hbar}{mc}$ [48]) intensifies the constant disjoint currents between each pair of adjacent QC particles in the G-group. However, the G-group resists the supersymmetry-breaking phenomenon. The resistance of the G-group primarily depends on the number and configuration of QC particles in the G-group as well as the amplitude of the external string fields.

Based on the vector form of Ohm's law, the conductivity of the G-group is correlated with the strength of the commutative SFIs of the Q-subgroup and the grand monopole in the G-group. According to the supersymmetry principle, the resistivity of the G-group is correlated to the strength of the noncommutative SFIs between the QC particles of the Q-subgroup and the G-group ($g_{el} \to g_{mag} \propto 1/g_{el}$ [26]).

As schematically shown in Figure 10, some of the Q-subgroup string fields ($m_\mu^Q$) and the string fields between QC particles in the G-group, $m_\mu^G$, arise as circular string fields around the G-group, which are known as magnetic fields in the literature. In contrast, the disjoint current between the Q-subgroup and G-group and the supersymmetry breaking of the G-group, which intensifies the disjoint currents between the QC particles of the G-group, depend on the resistivity of the G-group.

The undefined term of magnetic current density in Maxwell's equations has been proposed as the product of a coefficient and the magnetic field, with the coefficient represented under the nomenclature of magnetic conductivity versus electric conductivity [50]. Here, the magnetic current density corresponding to the disjoint current is proposed as the product of the resistivity of the G-group of QC particles and the circular string fields around the G-group, known as the magnetic fields in Maxwell's equations.

As the vector form of Ohm's law relates the macroscopic electric current density to an electric field, the macroscopic magnetic current density can be related to a magnetic field with the vector form of Ohm's law.

$$j = \sigma E \qquad (6.1a)$$
$$m = \rho H \qquad (6.1b)$$

where $\sigma$ and $\rho$ are the conductivity and resistivity of the G-group, respectively, and $m_\mu = \frac{m}{\rho}$.

The electric and magnetic currents, which are proportional to the conductivity and resistivity of the G-group, are correlated with the S-duality in the literature, in which the coupling constants are inverted as $g_{el} \to g_{mag} \propto 1/g_{el}$ [26]. Moreover, they are related to the background-independent canonical momentum components in (4.5a) and (4.5b).

As postulated in the context of relativity [51], the macroscopic magnetic and electric fields can be considered as two sides of the same phenomenon: these two macroscopic fields combined with their corresponding microscopic currents are two parts of the same phenomenon not only in the relativistic context (background-dependent) but also in the context of nonrelativistic (background-independent) SFIs.

## 7. Spin of gauge group's string fields and T-duality

As discussed in Section 3, the concept of the spin of an electron's string fields is associated with the noncommutative SFI of the electron and its adjacent electron (two adjacent QC particles). Due to the noncommutative SFI of the two QC particles, circular closed-loop string fields arise around the QC particles. In this section, the concept is generalized to a G-group of QC particles. The primary result of the spin of the G-group's string fields is the realization of quantized microscopic closed string fields corresponding to the macroscopic magnetic field in Maxwell's equations. These quantized microscopic closed string fields arise from the noncommutative SFI of QC particles associated with disjoint currents in the G-group. The secondary result of the spin of the G-group string fields is the demonstration of T-duality between the interior strong in-group SFIs of the G-group in a small volume and the closed string fields in a large volume around the G-group [52]. As explained in Section 5 [Figure 9, Equation (5.11)], a portion of the total disjoint current of the G-group, $m_\mu^G$, is translated to the next G-group, thus demonstrating both noncommutative and commutative SFIs ($\mathcal{J}_{GG}^\mu$) (see Figure 12). Another portion of the disjoint current demonstrates self-adjoint coupled string fields, $m_G^O$, in the shape of closed string fields. Figure 10 schematically shows the G-group under an infinitesimal gauge string field translation from the Q-subgroup. The infinitesimal variations in the disjoint currents contribute to the closed string fields around the G-group. Therefore, the same scenario for the string field spin of an electron can be considered for the string fields of the G-group on a macroscopic scale. The model represents the quantized form of the Ising model, known as magnetic dipole moments of atomic spin [53].

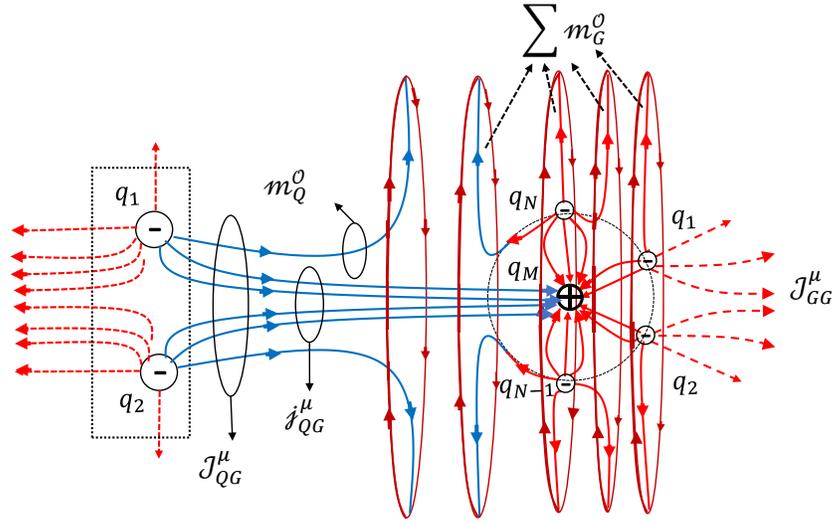

Figure 10. Spin of a G-group's String fields in atomic scale.

Figure 11(a) shows a vector diagram of the spin of the G-group string fields. Comparison of the spin of an electron string field in Figure 3 and that of a G-group in Figure 11 indicates that the spin of the G-group string fields is the summation of the string field spins of QC particles in the G-group. The G-group string field spin includes the two conjugated Dirac fields, whereas the electron string field spin merely includes one of the fields. The advanced characteristics of the G-group string field spin explain the undefined term of magnetic current as well as the electric current in Maxwell's equations, without the need for a permanent monopole. Moreover, the macroscopic model of the world tube diagram developed in string theory can be constructed to unify the quantum theory and general relativity in the anti-de Sitter space geometry (see Section 8.3). The G-group shown in Figure 9 is exemplified by four QC particles commuting with the grand monopole at the center. The 1/N expansion theory can be applied to the G-group with an infinite number of QC particles ($N \to \infty$). For a G-group with $N$ QC particles ($N \to \infty$), the derived equation for the microscopic disjoint current $m_\mu^G$ in (5.8) implies the generation of $N$ macroscopic circular closed-loop string fields separated by an infinitesimal distance. Therefore, an infinite number of quantized macroscopic closed-loop strings can be considered as a world tube, as shown in Figure 11(b).

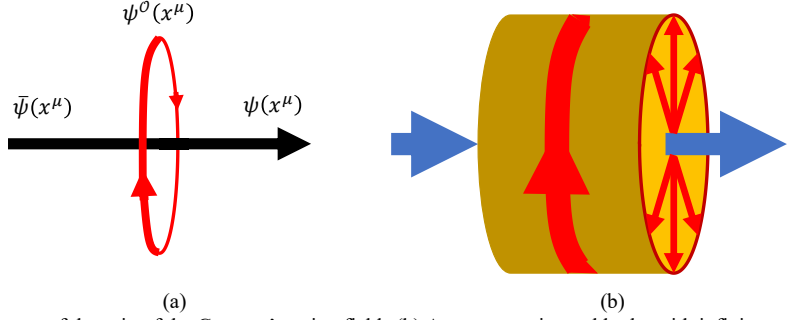

(a)              (b)

Figure 11. (a) Vector diagram of the spin of the G-group's string fields (b) A macroscopic world tube with infinite number of closed string fields.

## 8. Strong in-group and weak intergroup SFIs; S- and U-dualities
### 8.1. Intergroup commutative and noncommutative SFIs of entangled $G_2$ gauge groups

Thus far, commutative and noncommutative SFIs have been discussed for a G-group of QC particles. A locally supersymmetric spatial geometry was constructed for the G-group based on strong in-group commutative and noncommutative SFIs complying with abelian and non-abelian Yang-Mills string field theory.

In this section, gauge couplings between multiple G-groups of QC particles are explored based on weak intergroup commutative and noncommutative SFIs. The entangled G-groups of QC particles with both communitive and noncommutative SFIs are shown to break the local supersymmetry of each G-group and demonstrate a global supersymmetry for the spatial geometry of the entangled G-groups (homological mirror symmetry) [54].

A QC particle can generally realize strong in-group commutative and noncommutative SFIs with the grand monopole and other QC particles in the G-group as well as weak intergroup commutative and noncommutative SFIs with the grand monopole and QC particles of another G-group. Considering a G-group with $N$ QC particles and $M$ string fields associated with each QC particle, the SFIs of the particles can be mathematically represented as follows:

$N$ QC particles, including both QED and QCD particles ($N = N_{QC} = N_{QED} + N_{QCD}$), are assumed in a G-group (see Figure 7), in which the string fields of each particle are categorized as follows:

- $M_\alpha$, the quantum number of the string fields of a QC particle in the strong in-group commutative SFI with the grand monopole of the G-group.
- $M_\beta$, the quantum number of the string fields of the QC particle in the strong in-group noncommutative SFI with other adjacent QC particles in the G-group.
- $M_\gamma$, the quantum number of the string fields of the QC particle in weak intergroup SFIs. These string fields can be considered for the weak intergroup SFIs between two G-groups of QC particles and gravity $M_\gamma = M_{\gamma_i} + M_{\gamma_g}$ (known as Higgs field interactions in the literature [55]). Similarly, the string fields associated with the weak intergroup SFIs of the QC particle can be considered as commutative and noncommutative SFIs associated with adjoint and disjoint currents, respectively, $M_\gamma = M_\gamma^A + M_\gamma^D$. Thus, the total string fields of each particle are the summation of all three categories, $M_P = M_\alpha + M_\beta + M_\gamma$.

Moreover, the string fields of a grand monopole in a G-group can be considered as two classes, including those for SFIs with QC particles in the G-group or out of the G-group, with the gravitational field interactions, $M_M = M_\zeta + M_\chi$.

Figure 12 shows the topology of two entangled G-groups of QC particles, $G_I$ and $G_{II}$, represnted by the $G_2$-group. The weak intergroup gauge couplings of the $G_2$-group include both adjoint and disjoint couplings corresponding to commutative and noncommutative SFIs. As shown in Figure 12, the weak intergroup commutative and noncommutative SFIs of the $G_I$- and $G_{II}$-groups are represented as follows:

$Ad_{G_2}: G_I \leftrightarrow G_{II}$

$$\hat{\psi}_{G_{I,II}}(x^\mu) \equiv \left\langle \psi_{G_I^A}(x^\mu) \middle| \bar{\psi}_{G_{II}^A}(x^\mu) \right\rangle \tag{8.1a}$$

$$\hat{\psi}_{G_{II,I}}(x^\mu) \equiv \left\langle \psi_{G_{II}^A}(x^\mu) \middle| \bar{\psi}_{G_I^A}(x^\mu) \right\rangle \tag{8.1b}$$

Each pair of nonconjugated $\{\psi_{G_I^A}$ and $\psi_{G_{II}^A}\}$ and conjugated $\{\bar{\psi}_{G_I^A}$ and $\bar{\psi}_{G_{II}^A}\}$ string field functions represents the string fields of the QC particles and the grand monopoles in the two $G_I$- and $G_{II}$-groups in the weak intergroup commutative SFIs.

$$\psi_{G_I^A}(x^\mu) = \psi_{G_{II}^A}(x^\mu) = e^{-i\phi_{QG}} \sum_{n=1}^{N_\gamma^A} \sum_{m=1}^{M_\gamma^A} \varphi_n^m(x^\mu)\, e^{\pm j\phi_n^m(x^\mu)} \tag{8.2a}$$

$$\bar{\psi}_{G_I^A}(x^\mu) = \bar{\psi}_{G_{II}^A}(x^\mu) = e^{+i\phi_{MG}} \sum_{m=1}^{N_\gamma^A \times M_\gamma^A} \varphi_M^m(x^\mu)\, e^{\pm j\phi_M^m(x^\mu)} \tag{8.2b}$$

where $N_\gamma^A$ and $M_\gamma^A$ are the number of QC particles and their string fields in the commutative SFIs, respectively. $\phi_{QG}$ and $\phi_{MG}$ are the intrinsic background-independent phases of the subgroup of QC particles and the grand monopole of the G-groups.

$Di_{G_2}: G_I \leftrightarrow G_{II}$

$$\check{\psi}_{G_{I,II}}(x^\mu) \equiv \left(\psi_{G_I^D}(x^\mu)\middle|\psi_{G_{II}^D}(x^\mu)\right) \tag{8.3a}$$

$$\check{\psi}_{G_{II,I}}(x^\mu) \equiv \left(\psi_{G_{II}^D}(x^\mu)\middle|\psi_{G_I^D}(x^\mu)\right) \tag{8.3b}$$

The pair of nonconjugated $\{\psi_{G_I^D}$ and $\psi_{G_{II}^D}\}$ string field functions represent the string fields of the QC particles of the $G_I$- and $G_{II}$-groups in the weak intergroup noncommutative SFIs.

$$\psi_{G_I^D}(x^\mu) = \psi_{G_{II}^D}(x^\mu) = e^{-i\phi_{QG}} \sum_{n=1}^{N_\gamma^D} \sum_{m=1}^{M_\gamma^D} \varphi_n^m(x^\mu) e^{\pm j\phi_n^m(x^\mu)} \tag{8.4}$$

where $N_\gamma^D$ and $M_\gamma^D$ are the number of QC particles and their string fields in the noncommutative SFIs, respectively.

According to the commutative and noncommutative SFIs corresponding to the adjoint and disjoint gauge intergroup couplings, respectively, the SFIs of the entangled gauge groups of $G_I$ and $G_{II}$ ($G_2$-group) demonstrate a spatial geometry with (homological) mirror symmetry [54]. In addition to the intergroup SFIs, the in-group adjoint and disjoint currents associated with the commutative and noncommutative SFIs for each G-group can be derived in a similar manner for the entangled $G_2$-group.

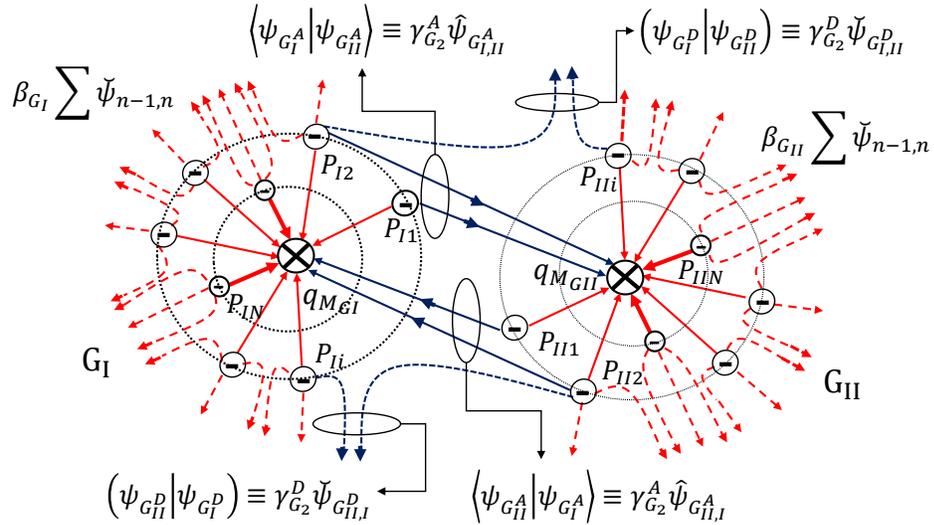

Figure 12. Weak intergroup commutative and noncommutative SFIs of the $G_I$ and $G_{II}$ gauge groups.

Similarly, the Hermitian and anti-Hermitian operators corresponding to the intergroup commutative and noncommutative SFIs are represented.

$$\langle \hat{A} \rangle \equiv \left\langle \psi_{G_I^A}(x^\mu) \middle| \hat{A} \middle| \bar{\psi}_{G_{II}^A}(x^\mu) \right\rangle \quad (8.5a)$$

$$\langle \hat{B} \rangle \equiv \left\langle \psi_{G_{II}^A}(x^\mu) \middle| \hat{B} \middle| \bar{\psi}_{G_I^A}(x^\mu) \right\rangle \quad (8.5b)$$

where

$$[\hat{A}, \hat{B}] = \hat{A}\hat{B} - \hat{B}\hat{A} = 0 \quad (8.6)$$

The physical interpretation of the relations is associated with the quantum potential energy of the entangled $G_2$-group in terms of commutative SFIs corresponding to Hermitian adjoint operators and can be observed (measured) through the quantum kinetic energy (see Figure A.3 in the Appendix). The relation between the anti-Hermitian disjoint operators can be similarly represented for the entangled $G_2$-group based on their noncommutative SFIs.

$$(\check{C}) \equiv \left( \psi_{G_I^D}(x^\mu) \middle| \check{C} \middle| \psi_{G_{II}^D}(x^\mu) \right) \quad (8.7a)$$

$$(\check{D}) \equiv \left( \psi_{G_{II}^D}(x^\mu) \middle| \check{D} \middle| \psi_{G_I^D}(x^\mu) \right) \quad (8.7b)$$

where

$$[\check{C}, \check{D}] = \check{C}\check{D} - \check{D}\check{C} = i\check{E} \quad (8.8)$$

Moreover, a wavefunction can be evolved forward in time by applying the time-evolution Hermitian and anti-Hermitian operators, e.g., for the commutative and noncommutative SFIs [38].

$$\langle \hat{A} \rangle \equiv \left\langle \psi_{G_I^A}(x,0) \middle| e^{i\hat{H}t/\hbar} \hat{A} e^{-i\hat{H}t/\hbar} \middle| \psi_{G_{II}^A}(x,0) \right\rangle \quad (8.9a)$$

$$(\check{C}) \equiv \left( \psi_{G_I^D}(x,0) \middle| e^{i\hat{H}t/\hbar} \check{C} e^{i\hat{H}t/\hbar} \middle| \psi_{G_{II}^D}(x,0) \right) \quad (8.9b)$$

where

$$|\psi_{G_I}(x,t)\rangle = e^{-i\hat{H}t/\hbar} \psi_{G_I}(x,0) \quad (8.10)$$

Inspection of the string field configuration of the entangled $G_2$-group with weak intergroup commutative and noncommutative SFIs verifies the S-duality between the strong and weak SFIs of the $G_I$- and $G_{II}$-gauge groups [56].

The vacuum energy density and/or cosmological constant are essentially related to gravity [57]. However, general relativity emphasizes the vital role of spatial geometry (spacetime) to describe gravity in terms of both attractive and repulsive forces on the cosmological scale [58]. An entangled G2-group not only compactifies commutative and noncommutative SFIs complying with abelian and non-abelian Yang-Mills field theory but also displays clear spatial geometry of a G-group of QC particles on a microscopic scale with interior confined SFIs. Here, the $G_2$-group is hypothetically proposed as the smallest component in the quantum vacuum, exhibiting the minimum energy interaction [$\check{E}$ in (8.8)] with gravity through the Higgs mechanism [55]. The entangled G2-gauge group can be considered as a dynamic vacuum that includes both commutative and noncommutative SFIs. The spontaneous supersymmetry breaking in the spatial geometry of the $G_2$-group can be explored according to the commutative and noncommutative SFIs with a supercharge of Q (see Figure A.3 in the Appendix).

The commutative and noncommutative SFIs in the spatial geometry of the entangled $G_2$-group can be categorized as in-group and intergroup SFIs. Both commutative and noncommutative in-group SFIs are stronger than their intergroup counterparts. Therefore, the commutative and noncommutative intergroup SFIs between the $G_I$- and $G_{II}$-gauge groups in the entangled $G_2$-group have been linked to the electroweak force in the literature [59]. Figure 13 shows the commutative and noncommutative SFIs of the entangled $G_2$-group with Q- and M-subgroups (see Figure A.2 for Q- and M-subgroups). Comparison of the given diagram with Feynman's diagram for electroweak SFIs [60] indicates that the latter merely represents the commutative SFIs complying with the abelian Yang-Mills string field theory. The proposed diagram represents both commutative and noncommutative SFIs between the $G_I$- and $G_{II}$-gauge groups complying with abelian and non-abelian Yang-Mills string field theory.

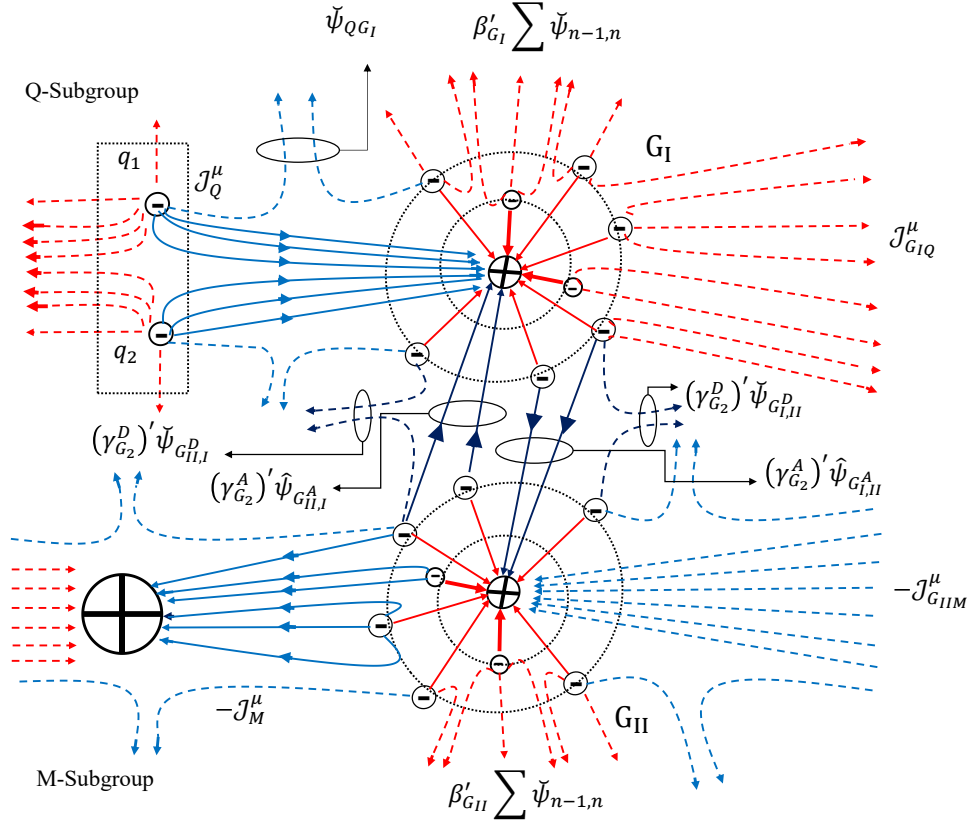

Figure 13. Commutative and noncommutative SFIs of the entangled $G_2$-group with two Q- and M-subgroups.

As shown in Figure 10, T-duality is verified for the interior strong SFIs in the G-group of QC particles and the exterior circular closed SFIs associated with the spin of the G-group string fields. Moreover, S-duality is illustrated for the strong in-group and weak intergroup SFIs between two entangled $G_2$-groups of QC particles in Figure 12. Consideration of the entangled $G_2$-group as the smallest component in the vacuum indicates that the mirror symmetry is broken as a result of SFIs with external strings fields, as shown in Figure 13. The coefficients of the in-group and intergroup commutative and noncommutative SFIs of the $G_2$-group change due to the symmetry breaking (see Section 8.3). The variations in the noncommutative SFI coefficients, $\beta_{G_I}$ and $\beta_{G_{II}}$, result in circular closed string fields around the entangled $G_2$-group (see Figure 14). U-duality can be verified as the unification of T- and S-dualities by the addition of circular closed string fields around the entangled $G_2$-group [61].

### 8.2. Rotating commutative and noncommutative vector string fields and torus construction

In Sections 3 and 6, self-adjoint SFIs were noted for both a QC particle and a G-group of QC particles. As posited, the string fields of two QC particles demonstrate noncommutative SFIs. If the two QC particles are taken to be identical, the same principle should apply to the string fields of a QC particle for self-adjoint SFIs. The noncommutative SFIs of two QC particles realize angular canonical momentum, which includes both Hermitian and anti-Hermitian operators. Figure 14 schematically illustrates a macroscopic circular spatial geometry realized by the rotating self-adjoint string fields shown in Figure 10. Here, the macroscopic rotating vector string fields are demonstrated by using an infinite number of G-groups of QC particles ($G_N - \text{group}, N \to \infty$).

Based on QCD screening and anti-screening principles [62], the noncommutative SFIs corresponding to the disjoint currents of a G-group polarize the infinite number of $G_2$-gauge groups in the vacuum to achieve self-adjoint coupling. The angular canonical momenta of the string fields produce a torus spacetime geometry by compactifying the commutative and noncommutative SFIs within and at the boundary of the torus [63].

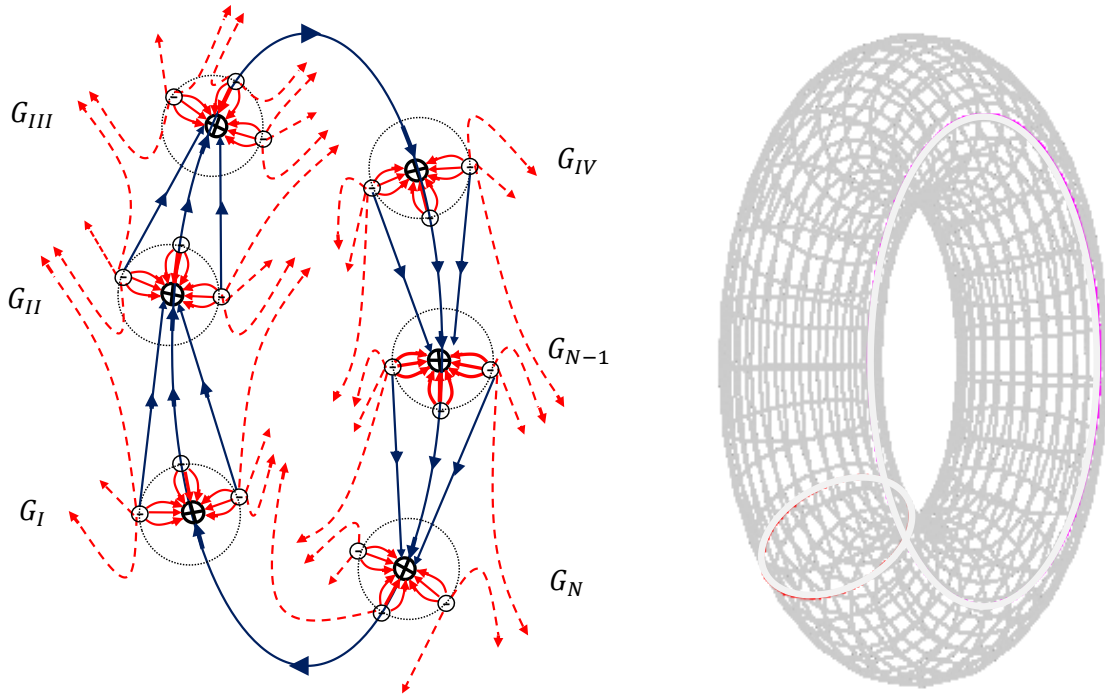

Figure 14. Macroscopic circular spatial geometry realized by the rotating self-adjoint SFIs of a G-group of QC particles (torus, the right picture from Wikipedia).

Inspection of the macroscopic circular spatial geometry of the $G_N$-group in Figure 14 indicates that each G-group demonstrates both commutative and noncommutative SFIs with the adjacent G-groups. The noncommutative SFIs between the G-groups make secondary macroscopic circular rotating vector string fields around each G-group in the interaction. In turn, the noncommutative SFIs of the G-groups in the secondary rotating vector string fields lead to tertiary macroscopic circular rotating vector string fields around each G-group in the secondary circular string fields, and so on. Figure 15 schematically shows a current-carrying wire in which macroscopic circular rotating vector string fields can be observed by measuring the electric currents in the loops. The spatial geometry in Figure 15 may be associated with a Calabi–Yau manifold [64].

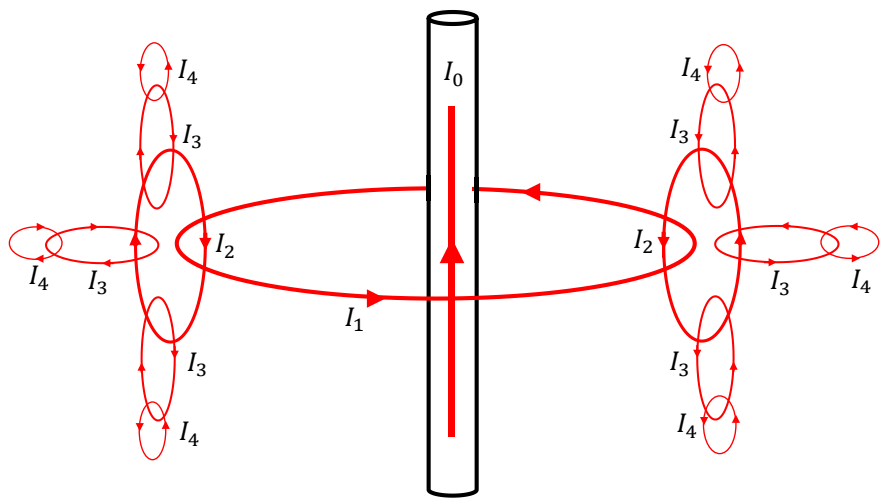

Figure 15. A current-carrying wire with infinite number of the macroscopic circular rotating vector string fields.

### 8.3. Background-independent (fixed) and background-dependent (dynamic) string field analyses

The background independence concept has recently been raised by emphasizing the structural foundation of quantum theory to resolve some underlying controversies in the literature [32]-[34]. Together with Newtonian mechanics, an ultimate theory based on the background independence concept and simultaneously complying with relativity and uncertainty principles would unify classical and modern physics [65]. This grand unification with the common concern of supersymmetry requires the reconciliation of quantum particle and quantum field theories, which historically stem from the concept of rigid particle and wave duality [66]. In contrast to quantum mechanics, which primarily emphasizes discrete point particles and/or their different combinations (hadrons), string theory describes the continuous characteristics of fields. Quantum mechanics focuses on the fundamental electromagnetic, weak, and strong forces associated with a set of massive and massless particles, whereas string theory focuses on gravitational forces based on commutative and noncommutative SFIs [67].

More generally, a main concern in contemporary physics is the search for an ultimate building block to unify and explain the underlying phenomena [11]-[13], [33]-[34] and [65]. This building block should include quantum particles, commutative and noncommutative SFIs in a single G-group of QC particles complying with Newtonian mechanics, and relativistic and non-relativistic perturbative string field theories. Moreover, the building block should unify and explain the relation between the ever-decaying radiative electromagnetic force and the permanent electroweak, strong, and gravitational forces.

Different aspects of an ultimate theory have been individually studied in the literature; however, the literature lacks a single building block to represent a consistent unification of the underlaying interactions. For example, commutative and noncommutative SFIs have been theoretically studied based on abelian and non-abelian Yang-Mills theory [9], but no study has compactified both noncommutative and commutative SFIs in a single G-group of QC particles as a building block. This building block should explain why spacetime partially commutes and partially not commute [10]. Moreover, the building block should describe the dynamic relations among the fundamental electromagnetic, strong, weak, and gravitational interactions in background-independent mechanics. These concerns have been partially discussed in the previous sections. In this section, the perturbative propagation of commutative and noncommutative string fields is dynamically illustrated in background-independent spacetime by using spontaneous diffeomorphism or supersymmetry breaking in the spatial geometry of a G-group of QC particles [68].

Figure 16(a) shows three individual G-groups of QC particles suspecting from an anchored bar and an instant current of zero-mode string fields specified by Dirac's delta function $[\Delta\psi_Q \delta(t)]$. This representation denotes the G-groups isolated from the permanent gravitational SFIs or the Higgs mechanism in the literature. Each G-group of QC particles can be replaced by an entangled $G_2$- or $G_N$-gauge group of QC particles to represent a quantum vacuum or a substance.

Based on the theory of general relativity in background-independent Newtonian mechanics, the exterior noncommutative SFIs, as the noncommutative boundary of the QC-particle G-group geometry, locally curve the vacuum spacetimes around each G-group of QC particles at their normal positions, thus complying with a permanent gravitational SFI [69]. In contrast, Maxwell's equations describe macroscopic electric and magnetic fields correlated to commutative and noncommutative SFIs of an infinite number of entangled G-groups of QC particles as a dipole magnet (Figure A.2) and a conductor (Figure A.2 with no broken supersymmetry) [1]. These equations can be reviewed for a microscopic QC-particle G-group or an entangled $G_2$-gauge group as a quantum vacuum. As shown in [10], the SFIs of a G-group of QC particles with external string fields are considered to evaluate Maxwell's equations in the microscopic scale.

After the barrier B shown in Figure 16(b) is removed, the string fields $\Delta\psi_Q \delta(t)$ and the open string fields of the $G_I$-group demonstrate both noncommutative and commutative SFIs at $t = t_1^-$ and $t = t_1^+$, respectively.

As discussed in Section 5, the spatial geometry of the $G_I$-group is spontaneously broken due to the commutative and noncommutative SFIs. At $t = t_2$, both the noncommutative and commutative SFIs are translated to the next $G_{II}$-group with corresponding canonical momenta $\breve{p} = -i\hbar\nabla\breve{T}$ and $\hat{p} = i\hbar\nabla\hat{T}$, respectively (see Section 4.2) [46]. By assigning the two coordinate systems $X_\mu$ and $X_\nu$ to the local positions of the $G_I$- and $G_{II}$-groups, the relativistic Lorentz invarince principle can be verified for the commutative and noncommutative SFIs. The relativistic principle has been studied for perturbative propagation of commutative SFIs through G-groups of QC particles, where the commutative string fields of $\Delta\hat{\psi}_{G_I G_{II}}$ are translated to the $G_{II}$-group as the local potential difference between the two G-groups. The changes in the local potential string fields of the $G_I$- and $G_{II}$-groups have been mathematically represented as an electromagnetic tensor [70].

$$\hat{F}_{\mu\nu} = (\partial_\mu A_\nu - \partial_\nu A_\mu) \tag{8.11}$$

The electromagnetic tensor represents the local commutative string field potential between the $G_I$- and $G_{II}$-groups ($\Delta\hat{\psi}_{G_IG_{II}}$).

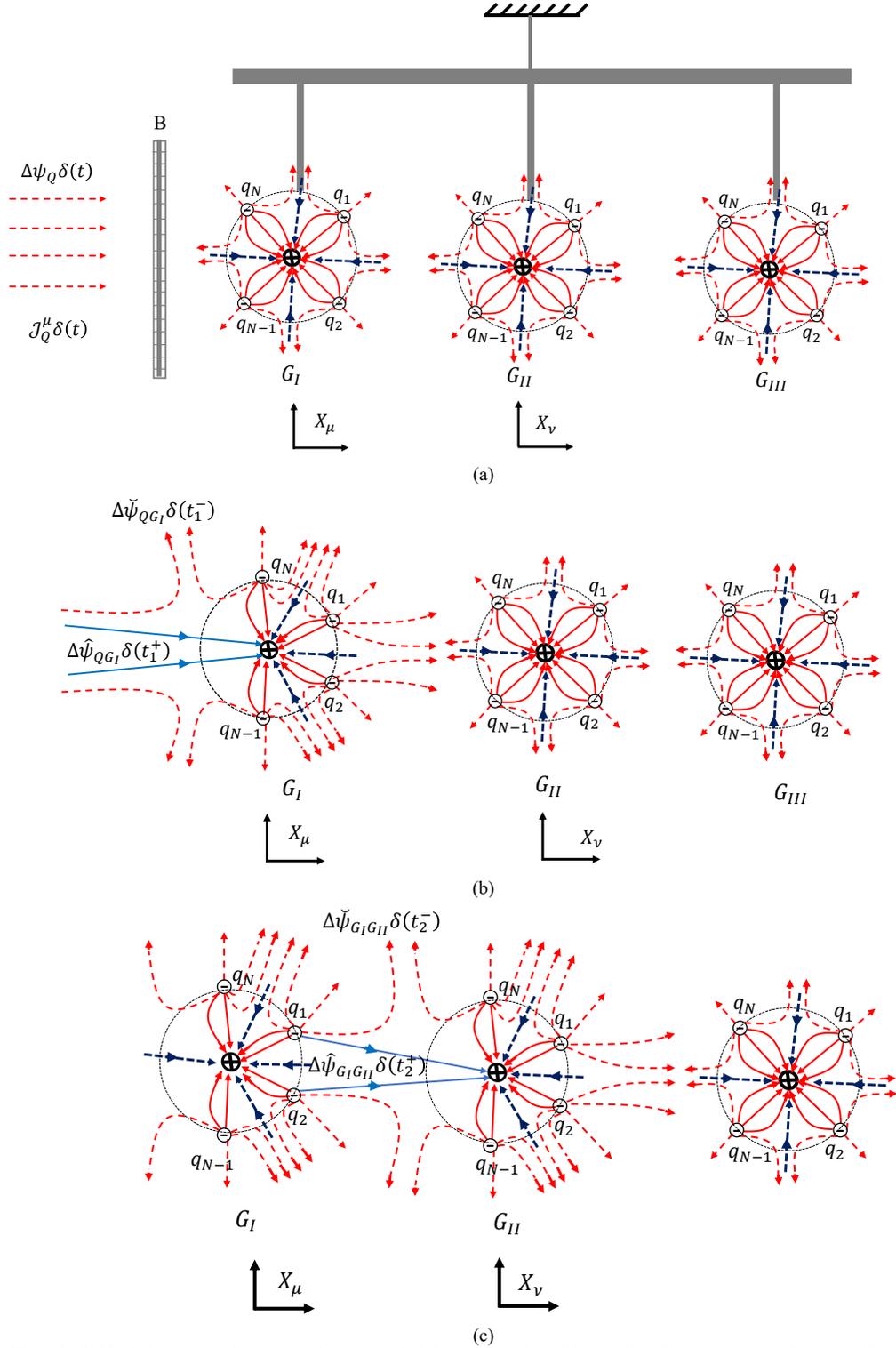

Figure 16. (a)Three individual G-groups of QC particles suspended from an anchored bar and an instant current of zero-mode string fields specified by Dirac delta function, translation of noncommutative and commutative SFIs from (b) Q-subgroup to $G_I$-group (c) $G_I$- group to $G_{II}$-group.

As discussed in Section 5, the translated supercurrent from the $G_I$-group demonstrates both commutative and noncommutative SFIs, corresponding to adjoint and disjoint currents. Following the variations in the local commutative string field potential ($\alpha_{G_{II}}\hat{\psi}_{G_{II}}$), the local noncommutative string field potential of the $G_{II}$-group ($\beta_{G_{II}}\breve{\psi}_{G_{II}}$) varies due to the noncommutative SFIs, with the translated supercurrent from the $G_I$-group. The variations in the noncommutative SFIs locally change the vacuum constant around the G-group of QC particles [71]. The variations in the noncommutative SFIs of the $G_{II}$-group can be represented based on infinitesimal gauge variations in the quantum number of the string fields translated from the $G_I$-group (see Section 5).

$$\partial^m \beta_{G_{II}} = -\partial^m \alpha_{G_{II}} + \partial^m \gamma_{G_I}^D \tag{8.12a}$$

and

$$\partial^m \alpha_{G_{II}} = -\partial^m \gamma_{G_I}^A \tag{8.12b}$$

where $\partial^m$ is the arithmetic derivative of the quantum number of the strings. $\alpha_{G_{II}}$ and $\beta_{G_{II}}$ are quantum coefficients of the in-group commutative and noncommutative SFIs of the $G_{II}$-group, and $\gamma_{G_I}^D$ is the intergroup noncommutative SFI of the $G_I$- and $G_{II}$-groups (see Section 5 and Figure). The arithmetic derivative of $\partial^m$ can be defined according to Fermat's quotient operator [72]. According to the definition of current density, the quantum number of the strings in the interactions is a function of spacetime, in which the arithmetic derivative of $\partial^m$ can be merged with the covariant spacetime derivative of $\partial_\mu$ as $\partial_\mu^m$ (see Figure 9 and Figure). Thus, the translation of noncommutative SFIs from one G-group to the next group can also be represented as a tensor.

$$\breve{F}^{\mu\nu} = (\partial^\mu M^\nu + \partial^\nu M^\mu) \tag{8.13}$$

The noncommutative and commutative SFI tensors provide a supersymmetric picture of the fundamental interactions. As shown in Figure 17(a), the instant commutative and noncommutative potentials are translated to the $G_{II}$-group, and the $G_I$-group attains its initial normal state. In the normal state, the squeezed QC particles of the G-group are relaxed, and the initially constant noncommutaive SFIs of the G-group and surronding spacetimes are retained. Figure 17(b) and (c) show the same translation for the commutative and noncommutative SFIs. According to the uncertainty and special relativity principles, an observer with a speed less than the speed of the SFIs cannot determine the instant positions of the canonical momentum components applied to the QC particles in the gauge groups with a given certainty [73].

In this work, commutative and noncommutative SFIs have been discussed primarily on the basis of the quantum number of the string fields. The quantum number of the string fields in a G-group of QC particles determines the spatial geometry of the G-group, in which the variations and translations of string fields mutually interact and influence the spacetime around the G-group of QC particles. By including the spacetime variations caused by commutative and noncommutative SFI variations, a comprehensive analysis of abelian and non-abelian string field theory can unify Newtonian mechanics and quantum electrodynamic and chromodynamic theories on the basis of a single (or multiple) background-independent G-group of QC particles that comply with the relativistic Lorentz invariance principle. Because the commutative SFIs and their translations are represented by the adjoint action of a Lie G-group of QC particles ($x \mapsto gxg^{-1}$) [74], the noncommutative SFIs and their translations should be differently denoted according to the disjoint action of a Lie G-group of QC particles ($x \mapsto gxg$) and matrix forms of the noncommutative actions.

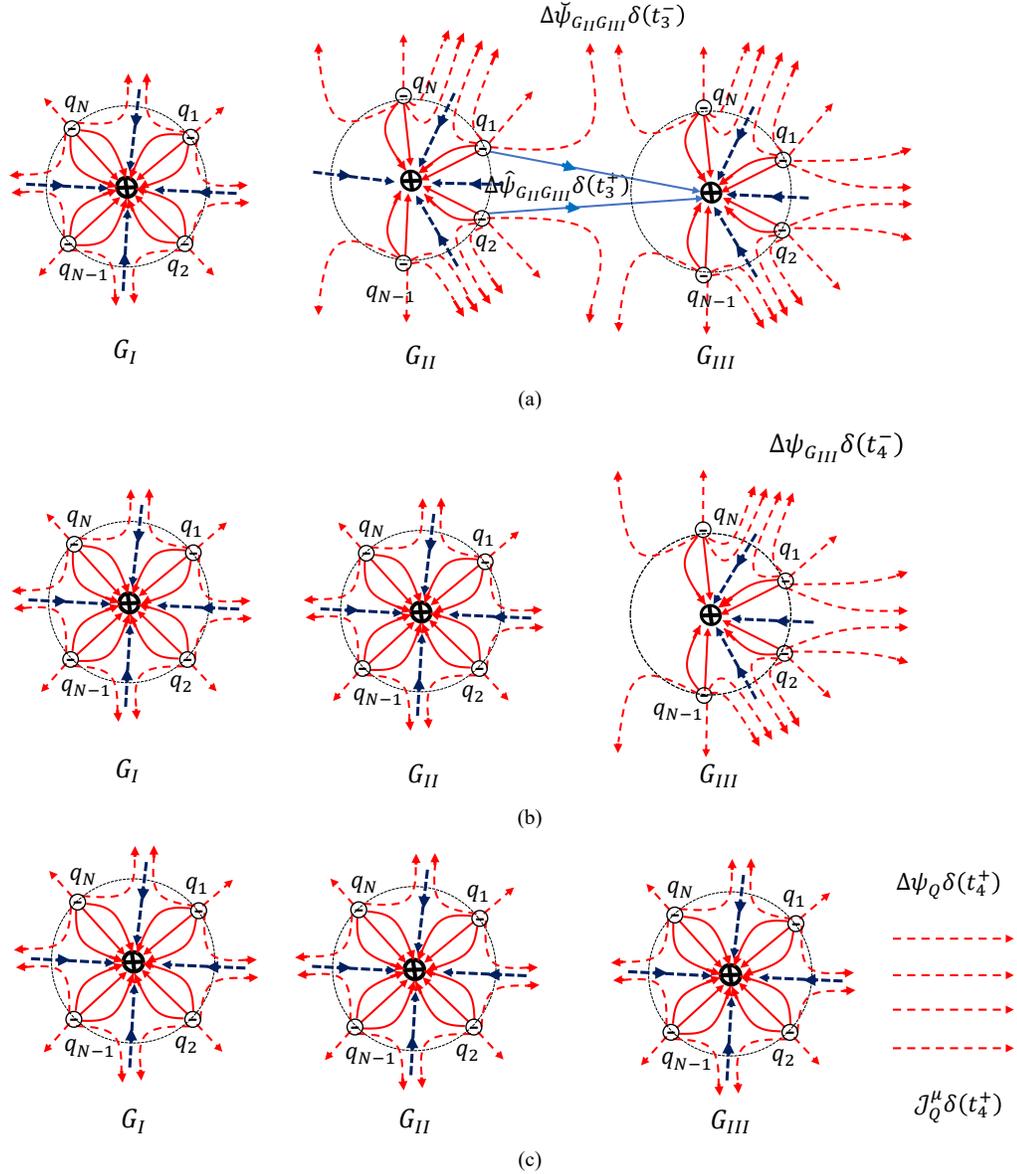

Figure 17. Translation of commutative and noncommutative SFIs from (a) $G_{II}$- group to $G_{III}$-group (b) $G_{III}$-group to load (c) the measured instant zero-mode string or current in load.

## 9. Conclusion

In this work, commutative and noncommutative SFIs have been incorporated in a single supersymmetric G-group of QC particles as an ultimate building block to unify the fundamental interactions. The building block explains the relativistic perturbative string field propagation in Newtonian background-independent mechanics. Based on in-group commutative and noncommutative SFIs, a supercurrent consisting of adjoint and disjoint currents was evaluated for a G-group of QC particles. Conjugated and nonconjugated string field functions were used to mathematically model the two currents based on Hermitian and anti-Hermitian operators. Because the microscopic adjoint current associated with the commutative SFI has already been related to the macroscopic electric current density in Maxwell's equations, it is proposed that the microscopic disjoint current associated with the noncommutative SFI is related to the undefined magnetic current density in these equations.

The vector form of Ohm's law relates the macroscopic electric current density to an electric field ($j = \sigma E$) with a coefficient of conductivity; similarly, it is proposed that the macroscopic magnetic current density is related to the macroscopic magnetic field on the basis of the vector form of Ohm's law ($m = \rho H$) with a coefficient of resistivity.

## Appendix

With the assumption of quanta $M$ ($M \to \infty$) string fields for a QC particle,

$$\psi_{\pm q}(x^\mu) = e^{\pm i\phi_q} \sum_{m=1}^{M} \varphi_q^m(x^\mu) e^{\pm j\phi_q^m(x^\mu)} \tag{A.1}$$

In the case of background-independent analysis and the assumption of quanta $m_K$ string fields of the particle in either commutative or noncommutative SFI:

$$\psi_{\pm q} = e^{\pm i\phi_q} \sum_{m=1}^{M} \varphi_q^m = e^{\pm i\phi_q} \left( \sum_{m=1}^{m_K} \varphi_q^m + \sum_{m=m_K+1}^{m_K} \varphi_q^m \right) \tag{A.2a}$$

$$\sum_{m=1}^{m_K} \varphi_q^m = \sum_{m=1}^{M} \varphi_q^m \left(1 - \frac{\sum_{m=m_K+1}^{m_K} \varphi_q^m}{\sum_{m=1}^{M} \varphi_q^m}\right) = \sum_{m=1}^{M} \varphi_q^m (1 - \zeta) = \xi \psi_{\pm q} \tag{A.2b}$$

where $\zeta$ is a real number smaller than one, the ratio of the quanta strings. The coefficient of $\xi$ is represented by one of the coefficients $\alpha$, $\beta$ or $\gamma$ for the in-group commutative, noncommutative, and intergroup SFIs, respectively. The coefficient $\gamma$ can be split to two coefficients of $\gamma_{ia}$ and $\gamma_{id}$ for intergroup commutative and noncommutative SFIS, respectively.

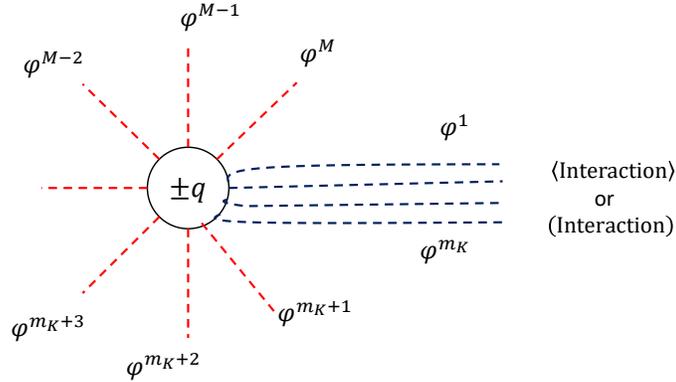

Figure A.1. Quanta number of the string fields of a charged particle in the interaction.

Under the assumption that $m_K$ zero-mode strings of $n_K$ QC particles of the $N$-QC-particle $G_N$-group in the atomic structure of the dipole magnet, shown in Figure A.2, contribute to the $\mathcal{J}_Q^\mu$ current, the string field function $\psi_Q$ associated with the north pole can be represented as (A.3a). In contrast, the string field function associated with the south pole of the dipole magnet can be represented as (A.3b).

$$\psi_Q(x^\mu) = e^{-i\phi_Q} \sum_{\substack{n=1 \\ n_K \times m_K}}^{n_K} \sum_{m=1}^{m_K} \varphi_n^m(x^\mu) e^{\pm j\phi_n^m(x^\mu)} \tag{A.3a}$$

$$\bar{\psi}_M(x^\mu) = e^{+i\phi_M} \sum_{m=1}^{n_K \times m_K} \varphi_M^m(x^\mu) e^{\pm j\phi_M^m(x^\mu)} \tag{A.3b}$$

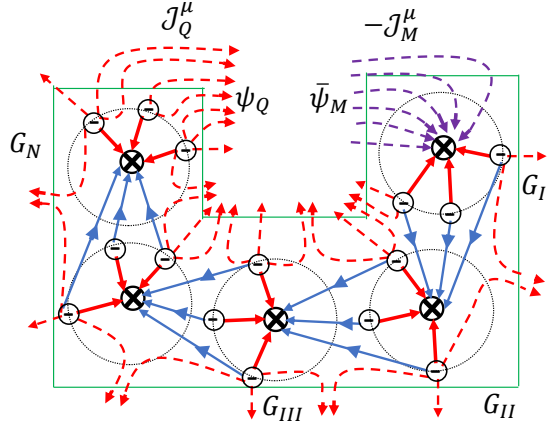

Figure A.2. Atomic structure of a typical magnet with the two conjugated field functions of $\psi_Q$ and $\bar{\psi}_M$.

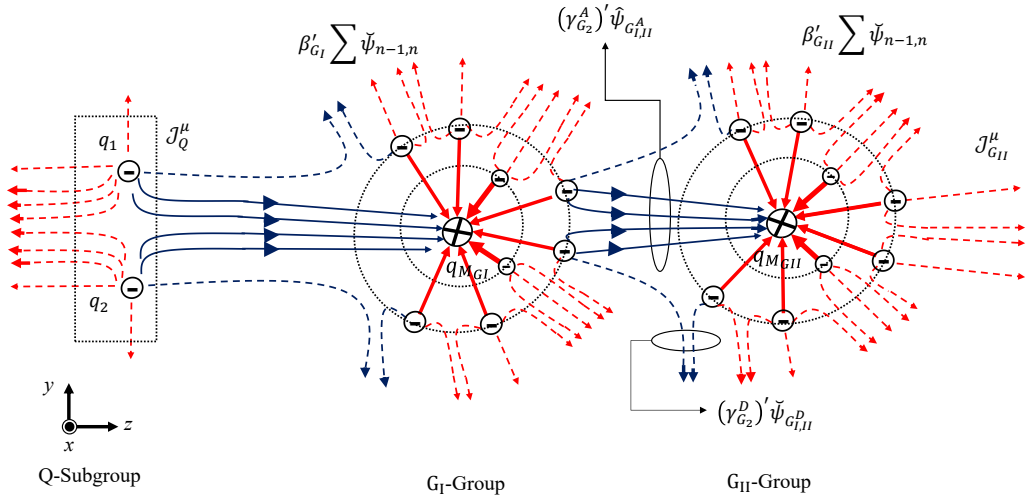

Figure A.3. Spontaneous supersymmetry breaking in the spatial geometry of $G_2$ gauge group.